\documentclass[prd,aps,floatfix,showpacs,tightenlines,superscriptaddress,amsmath,amssymb,showkeys,10pt,nofootinbib]{revtex4-2}

\usepackage{slashed}
\usepackage{xcolor}
\usepackage{graphicx}
\usepackage{float}
\usepackage{subcaption}
\usepackage{multirow}
\usepackage{hyperref}

\newcommand{\ket}[1]{  \left< #1 \right| }

\newcommand{\be}{\begin{equation}}\newcommand{\ee}{\end{equation}}
\newcommand{\bea}{\begin{eqnarray}}\newcommand{\eea}{\end{eqnarray}}
\newcommand{\brr}{\begin{array}}\newcommand{\err}{\end{array}}
\newcommand{\bit}{\begin{itemize}}\newcommand{\eit}{\end{itemize}}
\newcommand{\ben}{\begin{enumerate}}\newcommand{\een}{\end{enumerate}}

\newcommand{\bbm}{\begin{bmatrix}}\newcommand{\ebm}{\end{bmatrix}}
\newcommand{\ba}{\begin{array}}
\newcommand{\ea}{\end{array}}

\definecolor{darkred}{rgb}{.8,0,0}

\definecolor{darkblue}{rgb}{0,0,0}

\newtheorem{mydef}{Definition}
\newtheorem{Lemma}{Lemma}
\newcommand{\bd}{\begin{mydef}} \newcommand{\ed}{\end{mydef}}
\newcommand{\bthe}{\begin{theorem}} \newcommand{\ethe}{\end{theorem}}
\newcommand{\ble}{\begin{Lemma}} \newcommand{\ele}{\end{Lemma}}

\def\1{{_{1}}}\def\2{{_{2}}}

\def\noHe0{:\;\!\!\;\!\!:H_L(0):\;\!\!\;\!\!:}
\def\noHm0{:\;\!\!\;\!\!:H_\mu(0):\;\!\!\;\!\!:}

\def\1{{_{1}}}\def\2{{_{2}}}

\begin{document}

\title{Probing Quantum Foundations in Long-Baseline Neutrino 
Oscillations: \\ Wave Packet Approach from MINOS Data to DUNE 
Predictions}

\author{Baktiar Wasir Farooq}
\email{2330238@kmc.du.ac.in}
\affiliation{Department of Physics, Kirori Mal College, 
University of Delhi, India}

\author{Bipin Singh Koranga}
\email{bskoranga@kmc.du.ac.in}
\affiliation{Department of Physics, Kirori Mal College, 
University of Delhi, India}

\author{Massimo Blasone}
\email{mblasone@unisa.it}
\affiliation{Dipartimento di Fisica, Universit\`a di Salerno, Via Giovanni Paolo II, 132 84084 Fisciano, Italy \& INFN Sezione di Napoli, Gruppo collegato di Salerno, Italy}

\vspace{3mm}

\begin{abstract}
We investigate neutrino flavor oscillations in the wave packet formalism, extending the standard plane-wave treatment to incorporate the finite spatial coherence of neutrino mass eigenstates. We derive the full three-flavor oscillation probability with MSW matter effects, decoherence, and localization suppression governed by the wave packet width $\sigma_x$, benchmarking against MINOS disappearance data ($L = 735$ km, $\sigma_x \approx 0.24 \times 10^{-15}$ m) and extending to DUNE ($L = 1300$ km) predictions under the same parameters. To quantify how sharply the oscillation data constrain the wave packet width, we compute the classical Fisher information $F_C(\sigma_x, E)$ as a function of neutrino energy for both MINOS and DUNE, and show that its sensitivity is concentrated precisely in the low-energy regime where the wave-packet and plane-wave predictions diverge. Beyond the oscillation probability analysis, we compute wave–particle–entanglement complementarity quantities within a quantum information–theoretic framework and explore the complete triality description involving entanglement, predictability, and visibility --- as functions of neutrino energy for both experiments - MINOS and DUNE. We show that even when the oscillation probability is numerically indistinguishable from the plane-wave result, the wave packet structure leaves observable imprints in these triality measures. Our results demonstrate that the wave packet approach provides a theoretically consistent and experimentally motivated framework for probing quantum foundations in long-baseline atmospheric neutrino oscillation experiments.
\end{abstract}


 \maketitle

\section{Introduction}
\label{sec:intro}
 
Neutrino flavor oscillation remains one of the most direct experimental windows onto physics beyond the Standard Model, establishing that neutrinos are massive and that lepton flavor is not conserved during propagation \cite{Pontecorvo:1957}. In the standard three-flavor picture, flavor eigenstates $|\nu_\alpha\rangle$ are related to mass eigenstates $|\nu_j\rangle$ through the unitary PMNS mixing matrix, and the interference among the mass eigenstates as they propagate gives rise to the familiar $L/E$-dependent oscillation pattern. Long-baseline accelerator experiments have played a central role in pinning down the associated mixing parameters: MINOS, operating over a baseline of $L=735$~km, provided some of the most precise muon-neutrino disappearance measurements of the past decade \cite{Sousa:2015}, while the upcoming Deep Underground Neutrino Experiment (DUNE), with its considerably longer $L=1300$~km baseline, is expected to sharpen our knowledge of the mass ordering, the CP-violating phase $\delta_{\rm CP}$, and the mixing angles well beyond current global-fit precision \cite{Esteban2024NuFit60,NunokawaCP,EarthTomography}.
 
Almost all of this phenomenology is built on the plane-wave treatment of neutrino propagation, in which each mass eigenstate is idealized as an infinite, perfectly monochromatic wave. This idealization is analytically convenient, but it is not what is actually produced or detected in an experiment: neutrinos are created and absorbed in localized weak-interaction processes with finite energy-momentum uncertainty, so a more faithful description assigns each mass eigenstate a localized wave packet of finite spatial width $\sigma_x$ \cite{Giunti2004}. This quantum-mechanical refinement is not merely formal --- it introduces a finite coherence length beyond which the mass-eigenstate wave packets cease to overlap in space and time, causing the oscillation pattern to decohere and the transition probability to relax toward its flavor-averaged value, an effect that is entirely absent in the plane-wave limit. Wave-packet decoherence have been studied in a variety of contexts, from minimal-length-motivated modifications of the dispersion relation \cite{Ettefaghi:2024} to its impact on low-energy sterile-neutrino searches \cite{Arguelles2023}, and are expected to become increasingly relevant as baselines and detector distances grow, making them a natural and largely unexplored ingredient in wave-packet predictions for DUNE.

The wave-packet width $\sigma_x$ is not a universal constant but a property of the specific production and detection process, and its value must ultimately be fixed by experiment rather than by theoretical estimate alone. For MeV-scale reactor antineutrinos, in Ref.~\cite{deGouvea:2021ymm} it was performed a combined $\chi^2$ analysis of Daya Bay, RENO, and KamLAND data to place a data-driven lower bound, $\sigma > 2.1\times
10^{-4}$~nm at 90\% CL. Subsequently, in Ref.~\cite{Arg_elles_2023} it was shown that adopting this experimentally preferred value resolves the tension between reactor and gallium sterile-neutrino searches. We adopt the same data-driven approach here, but for a wave packet produced in pion/muon decay rather than nuclear beta decay:   we perform an analogous $\chi^2$ fit of $\sigma_x$ directly against MINOS $\nu_\mu$ disappearance data, rather than relying on the order-of-magnitude theoretical estimate $\sigma_x \sim \mathcal{O}(10^{-15})$~m of Ref.~\cite{Ettefaghi:2024}.
 
A recent line of research treats neutrino oscillation as a genuinely quantum-informational phenomenon rather than only a kinematic one. Because a flavor eigenstate produced via a charged-current interaction is a coherent superposition of the underlying mass eigenstates, its propagation can be recast as an entangling process among the flavor modes, and the resulting flavor eigenstate can be mapped onto a multi-qubit state amenable to the standard toolkit of quantum information theory~\cite{Blasone:2007wp,Blasone:2007vw,ill3,Blasone:2015lya,Blasone:2014jea,Alok:2014gya,Dixit:2018gjc,Song:2018bma,Ming:2020nyc,Jha:2020dav,Wang:2020vdm,Li:2021fft,Blasone:2021cau,Jha:2022yik,Bittencourt:2022tcl,Bittencourt:2023asd,Wang:2023rbf,ElBouzaidi:2025mwh}.

Complete complementarity relations~\cite{Jakob:2007zz,Jakob:2009gsy,Basso:2021whv} have shown to be a powerful tool both for investigating foundational aspects of quantum theory and for applications (see Ref.~\cite{Starke:2026iyr} for a review). In the context of neutrino oscillations, trade-off relations among quantum resources have been investigated in Refs.~\cite{Bittencourt:2022tcl,Li:2022zic,Bittencourt:2023asd}.
A similar approach has been followed in Ref.~\cite{Benerjee:2026}, based on the \emph{triality} among path predictability $\mathcal{P}^2$, interference visibility $\mathcal{V}^2$, and entanglement $\mathcal{E}^2$ satisfying $\mathcal{P}^2+\mathcal{V}^2+\mathcal{E}^2=1$ for a multipath quantum system \cite{PhysRevA.105.032209}.


Related quantum-information approaches --- including quantum and classical Fisher information analyses of oscillation parameters~\cite{Farooq:2026eap,yadav2026quantumfisherinformationrevealing,chundawat2026leptoniccpphasedetermination,Huang:2026bws,frugiuele2026quantumestimationtheorylimits, Lu:2026wzr}, studies of flavor-mode entanglement dynamics~\cite{Alok:2025qqr}, and broader reviews of quantum information as a precision probe of neutrino physics~\cite{Dixit:2026yft} --- underscore a growing consensus that entanglement and coherence measures encode oscillation physics not fully captured by the transition probability alone. Building on our own earlier work applying entanglement measures to two-flavor oscillations in matter \cite{SinghKoranga:2024rum} and quantum Fisher information to sterile-neutrino searches \cite{Farooq:2026eap}, we consider the question: how does the finite coherence of a wave packet, rather than an idealized plane wave, imprint itself on the triality decomposition of a propagating neutrino?
 
This question is not academic. Wave-packet decoherence and quantum-informational quantities such as $\mathcal{P}^2$, $\mathcal{V}^2$, and $\mathcal{E}^2$ are both sensitive to the same underlying coherence properties of the propagating state, yet they need not track the oscillation probability in the same way: a wave-packet correction can leave $P_{\mu\mu}(E)$ numerically indistinguishable from its plane-wave counterpart while still leaving a measurable imprint on the triality quantities, since the latter depend on the full density-matrix structure rather than a single diagonal element. Establishing whether and where such imprints appear therefore constitutes an independent, information-theoretic probe of the wave-packet formalism, complementary to direct fits of the oscillation probability itself.
 
In this paper, we address this question in four steps. First,  we derive the full three-flavor wave-packet oscillation probability, including Mikheyev--Smirnov--Wolfenstein (MSW) matter effects \cite{NunokawaCP} and the Earth's density profile \cite{dziewonski1981preliminary}, starting from a Gaussian momentum-space wave function for each mass eigenstate. Second, we confront this formalism with muon-neutrino disappearance data from MINOS at $L=735$~km, and show that a wave-packet width of $\sigma_x\approx0.24\times10^{-15}$~m yields improved qualitative agreement with the low-energy data relative to the plane-wave approximation; we then propagate the same formalism, with the constrained and independently varied values of $\sigma_x$, to generate wave-packet predictions for DUNE at $L=1300$~km, including its sensitivity to the neutrino mass ordering. Third, we use the classical Fisher information $F_C(\sigma_x,E)$ to quantify how sharply the oscillation probability constrains $\sigma_x$ as a function of energy for both experiments, and show that this sensitivity is concentrated precisely in the energy regime where the wave-packet and plane-wave predictions diverge. Finally, recasting the propagating neutrino as a three-qubit flavor state, we compute the triality quantities $\mathcal{P}^2$, $\mathcal{V}^2$, and $\mathcal{E}^2$ as functions of neutrino energy for both MINOS and DUNE, verify the triality identity $\mathcal{P}^2+\mathcal{V}^2+\mathcal{E}^2=1$ within the wave-packet framework, and identify the energy regions where wave-packet decoherence leaves an observable signature in these quantum-informational quantities even when the oscillation probability itself does not. Taken together, these results indicate that the wave-packet formalism, combined with quantum-informational diagnostics, provides a theoretically consistent and experimentally motivated framework for probing quantum foundations in long-baseline neutrino oscillation experiments.

The paper is organized as follows. In Section \ref{sec:theory}, we introduce the theoretical framework for the wavepacket formalism of neutrino oscillations including matter effects. In Section \ref{Minos Wavepacket section}, we apply the wavepacket formalism to the MINOS experiment data and work on a $\chi^2$  analysis to determine the wavepacket width $\sigma_{x}$, followed by a Jackknife Test to understand the neutrino energy sensitivity bins. In Section \ref{IV}, we perform theoretical predictions for the DUNE experiment  using the varying matter effects of the Earth. In Section \ref{CFI}, we study the sensitivity of  $\sigma_{x}$ by varying the Neutrino Energy (E) for both the experiments - MINOS and DUNE and  determine the most sensitive energy bins to wavepacket approach. In Section \ref{VI}, we perform a detailed analysis of the triality relations for both MINOS and DUNE. We summarize our findings and present some concluding remarks in Section \ref{sec:conclusion}.

\section{Theoretical Framework}\label{sec:theory}
The standard treatment of neutrino oscillations relies on the plane wave 
approximation, in which mass eigenstates are assumed to be completely 
delocalized in space. While this approach successfully captures the 
leading-order oscillation phenomenology, it fails to account for the 
finite spatial extent of the neutrino production and detection processes. 
The wave packet formalism addresses this limitation by assigning a 
localized wave function $\psi_j(x,t)$ to each mass eigenstate $|\nu_j\rangle$, 
characterized by a finite spatial width $\sigma_x$. This naturally 
introduces a coherence length $L^{\text{coh}}_{kj}$ — the maximum 
propagation distance over which the mass eigenstate wave packets remain 
spatially overlapping and can therefore interfere quantum mechanically. 
When $L \gtrsim L^{\text{coh}}_{kj}$, the wave packets separate and 
oscillations are exponentially suppressed, a feature entirely absent in 
the plane wave treatment. In what follows, we derive the full three-flavor 
wave packet oscillation probability including matter effects, starting 
from a Gaussian momentum distribution for each mass eigenstate.
\subsection{Flavor State as a Superposition of Wave Packets}
\label{subsec:flavor}
When we consider the plane wave approximation for the case of neutrinos, we assume that the mass eigenstate wavepackets are delocalized whereas for the case of wavepacket approach, we consider that the neutrino states are localized and for that an additional term $\psi_k$ is added in the standard notation.
A neutrino produced with flavor $\alpha$ at the 
space-time origin is described in the quantum mechanical 
wave packet formalism by the state
\begin{equation}
    |\nu_\alpha(x,t)\rangle = \sum_j U^*_{\alpha j}\,
    \psi_j(x,t)\,|\nu_j\rangle,
    \label{eq:flavor_state}
\end{equation}
where $U_{\alpha j}$ are elements of the 
Pontecorvo--Maki--Nakagawa--Sakata (PMNS) mixing 
matrix~\cite{Pontecorvo:1957}, $|\nu_j\rangle$ 
denotes the mass eigenstate with mass $m_j$, and 
$\psi_j(x,t)$ is the corresponding wave function.  
The standard parameterization of the PMNS matrix is
\begin{equation}
    U = \begin{pmatrix}
        c_{12}c_{13} & 
        s_{12}c_{13} & 
        s_{13}e^{-i\delta_{\rm CP}} \\[5pt]
        -s_{12}c_{23} - c_{12}s_{23}s_{13}
        e^{i\delta_{\rm CP}} &
        \phantom{-}c_{12}c_{23} - s_{12}s_{23}s_{13}
        e^{i\delta_{\rm CP}} &
        s_{23}c_{13} \\[5pt]
        \phantom{-}s_{12}s_{23} - c_{12}c_{23}s_{13}
        e^{i\delta_{\rm CP}} &
        -c_{12}s_{23} - s_{12}c_{23}s_{13}
        e^{i\delta_{\rm CP}} &
        c_{23}c_{13}
    \end{pmatrix},
    \label{eq:PMNS}
\end{equation}
where $c_{ij} \equiv \cos\theta_{ij}$, $s_{ij} \equiv 
\sin\theta_{ij}$, and $\delta_{\rm CP}$ is the 
CP-violating phase.

The momentum-space wave function of each mass eigenstate 
is taken to be a Gaussian of width $\sigma_p$ centered 
at the mean momentum $p_j$:
\begin{equation}
    \psi_j(p) = \left(2\pi\sigma_p^2\right)^{-1/4}
    \exp\!\left[-\frac{(p-p_j)^2}{4\sigma_p^2}\right].
    \label{eq:momentum_wp}
\end{equation}
The width $\sigma_p$ is determined by the momentum 
uncertainty of the production process. On further solving using the wavepacket logic as shown in the Appendix \ref{Appendix}, the final oscillation probability is expressed by :

\begin{equation}
    P_{\nu_\alpha \to \nu_\beta}(L) =
    \sum_{k,j} U^*_{\alpha k}\,U_{\alpha j}\,
    U_{\beta k}\,U^*_{\beta j}\;
    \exp\!\left[-2\pi i\frac{L}{L^{\rm osc}_{kj}}
    - \left(\frac{L}{L^{\rm coh}_{kj}}
    \right)^{\!2}\right].
    \label{eq:prob_xi0}
\end{equation}

\subsection{Matter Effects}
\label{subsec:matter}

Neutrinos propagating through matter experience a 
flavor-dependent potential from coherent forward 
scattering. The neutral-current contribution is 
flavor-universal and absorbed into an overall phase; 
the physical matter potential is
\begin{equation}
    V_{cc} = \sqrt{2}\,G_F\,n_e
    \approx 7.56 \times 10^{-14}\,Y_e\,
    \left(\frac{\rho}{{\rm g\,cm}^{-3}}\right)
    \,{\rm eV},
    \label{eq:matter_potential}
\end{equation}
where $G_F$ is the Fermi constant, $n_e$ is the 
electron number density, $Y_e$ is the electron 
fraction, and $\rho$ is the matter density, it's value varies depending on the experiments we consider, we prefer the PREM Model \cite{dziewonski1981preliminary} to parameterize the value of $\rho$ which is a function of the baseline length. The 
full Hamiltonian in the flavor basis is
\begin{equation}
H =
U
\begin{pmatrix}
0 & 0 & 0 \\
0 & \dfrac{\Delta m_{21}^{2}}{2E} & 0 \\
0 & 0 & \dfrac{\Delta m_{31}^{2}}{2E}
\end{pmatrix}
U^\dagger
+
\begin{pmatrix}
V_{cc} & 0 & 0 \\
0 & 0 & 0 \\
0 & 0 & 0
\end{pmatrix}.
\label{eq:Hamiltonian}
\end{equation}
which we diagonalize numerically at each energy:
\begin{equation}
H=W\begin{pmatrix}
\lambda_1 & 0 & 0 \\
0 & \lambda_2 & 0 \\
0 & 0 & \lambda_3
\end{pmatrix}
W^\dagger.
\end{equation}

Here, W and $\lambda_i$ ($i = 1,2,3$) are the PMNS matrix  and eigenvalues modified with 
matter. The wave packet oscillation probability in matter 
is obtained from (\ref{eq:prob_full}) by replacing 
$U \to W$ and $\Delta m^2_{kj}/2E \to 
\Delta\lambda_{kj} \equiv \lambda_k - \lambda_j$:
\begin{equation}
    P_{\nu_\alpha \to \nu_\beta}(L) =
    \sum_{k,j} W^*_{\alpha k}\,W_{\alpha j}\,
    W_{\beta k}\,W^*_{\beta j}\;
    \exp\!\left[-i2\pi \frac{L_{\rm iEV}}{L_{kj}^{osc,m}}
    - \left(\frac{L_{\rm iEV}}
    {L^{\rm coh,m}_{kj}}\right)^{\!2}- 2\pi^2\xi_p^2\left(\frac{\sigma_{iEV}}{L_{kj}^{osc,m}}\right)^2\right],
    \label{eq:prob_matter}
\end{equation}
where $L_{\rm iEV} = L \times 5.06773 \times 
10^9\,{\rm eV}^{-1}$, $\sigma_{iEV}$, $E$ and $\Delta\lambda_{kj}$ are in eV and
\begin{equation}
    L^{\rm coh,m}_{kj} = \frac{2\sqrt{2}\,
    \sigma_{\rm iEV} E}{|\Delta\lambda_{kj}|}
    \label{eq:Lcoh_matter}
\end{equation}
\begin{equation}
    L^{\rm osc,m}_{kj} = \frac{2\,
    \pi}{\Delta\lambda_{kj}}
    \label{eq:Losc_matter}
\end{equation}
are the coherence and the oscillation lengths in matter, respectively. 

\section{MINOS and Wavepacket Model}\label{Minos Wavepacket section}
In this section, we consider the datasets for MINOS, MINOS+ analyzed   in Ref.~\cite{Sousa:2015}. We start our analysis with the plane wave approximation and then carry forward the analysis with the wave packet approximation.The \textbf{M}ain \textbf{I}njector \textbf{N}eutrino \textbf{O}scillation \textbf{S}earch (\textbf{MINOS}) experiment is a long-baseline neutrino oscillation experiment using the Neutrinos at the Main Injector (NuMI) neutrino beam and two detectors to make precise measurements of neutrino oscillation parameters over a distance of 735 km. It aims to determine the oscillation of muon neutrinos into tau and electron neutrinos or sterile neutrinos, in our case we have limited ourself with electron and muon. Since September 2013, the MINOS experiment has become MINOS+, operating with the same detectors with upgraded electronics and the NuMI beam updated to the running conditions to be used during the NOvA era.
\vspace{0.5cm}
\begin{table}[h] \label{Table}
\centering
\begin{tabular}{|c|c|c|c|c|c|c|}
\hline
$\Delta m_{21}^2$ &
$\Delta m_{31}^2$ &
$\theta_{12}$ &
$\theta_{13}$ &
$\theta_{23}$ &
$L$ &
$\rho$\\
\hline
$7.39 \times 10^{-5}\,\mathrm{eV}^2$ &
$2.45 \times 10^{-3}\,\mathrm{eV}^2$ &
$33.82^\circ$ &
$8.61^\circ$ &
$49.7^\circ$ &
$735$ km &
$2.6$ $g/cm^3$\\
\hline
\end{tabular}
\caption{Neutrino oscillation parameters for MINOS \cite{Ettefaghi:2024}.}
\label{tab:MINOS_parameters}
\end{table}

\begin{figure}[t]
    \centering
\includegraphics[width=0.7\textwidth]{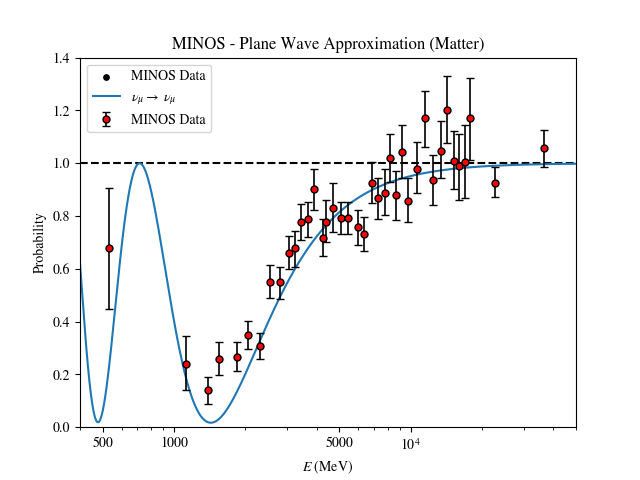}
    \caption{Three-flavor survival probability $\nu_{\mu} \rightarrow \nu_{\mu}$  in the plane wave approximation.}
    \label{fig:plane MINOS}
\end{figure}

The parameters we used for our probability plots for the case of MINOS are listed in Table 1. Here, the earth matter density $\rho = 2.6g/cm^3$ is constant all throughout because the maximum dept gained by the detector is $\approx$ 10.6 km, and the PREM model~\cite{dziewonski1981preliminary} predicts negligible density variation over this depth range, justifying the approximation of constant matter density for this detector. Hence, we consider constant matter density for this case. We further use these parameters to frame the atmospheric neutrinos survival probability for the MINOS by using the plane wave formalism and try to analyze our results with respect to the datasets.  

We can  observe from Fig.~\ref{fig:plane MINOS} that for the lower energy range $\approx (525 - 1500)$ MeV, the plane wave approximation doesn't appear to fit well with the datasets. Datasets for energy range $>1500$ MeV fits well. So, the current plane wave model have some flaws, specially within the smaller energy ranges, which must be fixed. We tried doing the same by using the wavepacket approximation and preferably got quite interesting results, the plot fits with the lower energy region and remains almost the same for higher energy region, see Fig.~\ref{fig:wave MINOS}. 

\subsection{Chi-Square Determination of the Wave-Packet Width}
\label{sec:chi2-fit}
To place the wave-packet width $\sigma_x$ on a quantitative, data-driven
footing rather than an order-of-magnitude estimate alone, we perform a
least-squares fit of $\sigma_x$ against the digitized MINOS $\nu_\mu$
disappearance data of Fig.~\ref{fig:wave MINOS}. Following the
statistical-errors-only approach of Ref.~\cite{Chan_2016}, we construct

\begin{equation}
\chi^2(\sigma_x) \;=\; \sum_{i=1}^{N}
\frac{\left[P_{\mu\mu}(E_i;\sigma_x) - P_i^{\rm data}\right]^2}
{\left(\sigma_i^{\pm}\right)^2},
\qquad
\sigma_i^{\pm} =
\begin{cases}
\sigma_i^{+} & \text{if } P_{\mu\mu}(E_i;\sigma_x) > P_i^{\rm data} \\
\sigma_i^{-} & \text{if } P_{\mu\mu}(E_i;\sigma_x) < P_i^{\rm data}
\end{cases}
\label{eq:chi2-def}
\end{equation}
where $P_{\mu\mu}(E_i;\sigma_x)$ is the wave-packet survival probability
of Eq.~\eqref{eq:prob_matter}, evaluated at the $N=39$ digitized MINOS
energy bins spanning $E\in[500,2\times10^4]$~MeV, and $P_i^{\rm data}$,
$\sigma_i^{\pm}$ are the corresponding central values and asymmetric
error bars read from Fig.~\ref{fig:wave MINOS}. The asymmetric-error
convention accounts for the fact that $P_i$ is derived from a ratio of
event counts and is not exactly Gaussian-distributed, particularly in
the low-statistics, low-energy bins.

\begin{figure}[t]
\centering
\includegraphics[width=0.7\textwidth]{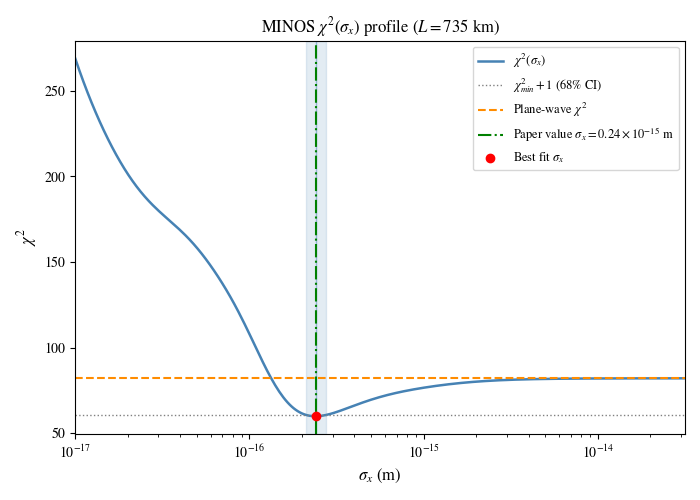}
\caption{The $\chi^2(\sigma_x)$ profile obtained from
Eq.~\eqref{eq:chi2-def} using the digitized MINOS data of
Fig.~\ref{fig:wave MINOS}. The dotted grey line marks
$\chi^2_{\min}+1$, defining the $68\%$ confidence interval (shaded
band); the dashed orange line is the fixed plane-wave $\chi^2$
($\sigma_x\to\infty$, zero free parameters); the red point marks the
best-fit $\sigma_x$, adopted throughout this work.}
\label{fig:chi2-profile}
\end{figure}

The resulting $\chi^2(\sigma_x)$ profile, obtained by scanning
$\sigma_x \in [10^{-17},10^{-13.5}]$~m, is shown in
Fig.~\ref{fig:chi2-profile}. The profile has the expected two-regime
structure: at small $\sigma_x$, the coherence length is too short even
for the lowest-energy bins, driving $\chi^2$ to large values as the
model over-suppresses oscillations across the whole spectrum; at large
$\sigma_x$, the damping term becomes negligible and $\chi^2$ asymptotes
to the fixed plane-wave value. Between these two regimes lies a well
defined minimum.

Minimizing Eq.~\eqref{eq:chi2-def} yields a best-fit value of
\begin{equation}
\sigma_x = 0.24\times10^{-15}\,\text{m},
\qquad
\chi^2_{\min} = 59.90,
\qquad
\chi^2_{\min}/\text{dof} = 1.58 \;\; (\text{dof} = N-1 = 38),
\end{equation}

with a $68\%$ confidence interval, obtained from the standard
$\Delta\chi^2=1$ criterion for a single free parameter (shaded band in
Fig.~\ref{fig:chi2-profile}), of
$\sigma_x \in [0.21,0.28]\times10^{-15}$~m. This value is consistent in
order of magnitude with the theoretical estimate of
Ref.~\cite{Ettefaghi:2024}, $\sigma_x \sim \mathcal{O}(10^{-15})$~m, and
is adopted as the wave-packet width used throughout this work, including
the DUNE predictions.

To quantify the statistical preference for the wave-packet treatment
over the standard plane-wave approximation, we compare $\chi^2_{\min}$
against the fixed (zero-free-parameter) plane-wave prediction,
$\sigma_x\to\infty$, shown as the dashed line in
Fig.~\ref{fig:chi2-profile}:
\begin{equation}
\chi^2_{\rm PW} = 82.06,
\qquad
\Delta\chi^2 \equiv \chi^2_{\rm PW} - \chi^2_{\min} = 22.16.
\end{equation}

Since the plane-wave limit is nested within the wave-packet model (it is
recovered by removing the single parameter $\sigma_x$), $\Delta\chi^2$
is approximately $\chi^2$-distributed with one degree of freedom by
Wilks' theorem~\cite{Wilks:1938dza}, corresponding to a preference for the
wave-packet treatment of $\sqrt{\Delta\chi^2}\approx4.7\sigma$. This
provides a quantitative statistical basis for the qualitative
improvement over the plane-wave fit already evident in
Fig.~\ref{fig:plane MINOS} at low energy.

We note that this fit treats the $39$ MINOS bins as statistically
independent, following the precedent of Ref.~\cite{Chan_2016}; the full
systematic covariance of the MINOS/MINOS+ measurement~\cite{Sousa:2015}
is not publicly tabulated and is therefore not included, which likely
results in a mild underestimate of the true uncertainty on $\sigma_x$.

\subsection{Jackknife Test of the Wave-Packet Preference}

The comparison in Eq.~(13) gives $\Delta\chi^2 = 22.16$ between the
plane-wave limit and the best-fit wave packet. To test how robust this
preference is, we performed a jackknife study on the $N=39$ energy bins.
In each case the wave-packet width $\sigma_x$ is refit using only the
retained bins, and $\Delta\chi^2 \equiv \chi^2_{\rm PW}-\chi^2_{\rm min}$
is recomputed.

\textit{Leave-one-out.} Removing any single bin leaves
$\Delta\chi^2 \ge 12.4$. The most influential bin is at
$E\simeq1551$~MeV, which alone contributes $\Delta\chi^2=10.7$, about half
of the total. Removing it gives $\Delta\chi^2=12.4$, and the best-fit
width shifts only from $2.4$ to $2.9\times10^{-16}$~m. Over all
leave-one-out fits the best-fit $\sigma_x$ stays within
$1.6$--$2.9\times10^{-16}$~m. The jackknife standard error on
$\Delta\chi^2$ is $10.9$, about half of its nominal value.

\begin{figure}[t]
  \centering
  \includegraphics[width=0.8\columnwidth]{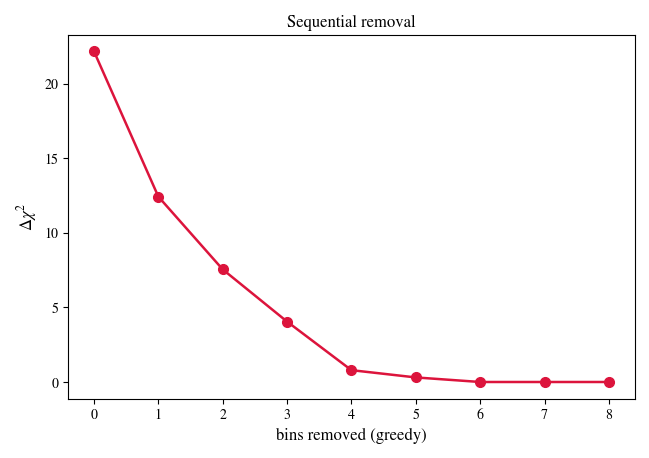}
  \caption{Jackknife test of the preference for a finite wave-packet
  width. $\Delta\chi^2=\chi^2_{\rm PW}-\chi^2_{\rm min}$ is recomputed,
  with $\sigma_x$ refit, after sequentially removing the bin whose removal
  lowers it most. The five bins removed first, at $E\simeq1551$, 1391,
  528, 1837 and 2059~MeV, account for the entire preference.}
  \label{fig:jackknife}
\end{figure}

\begin{figure}[t]
    \centering
\includegraphics[width=0.7\textwidth]{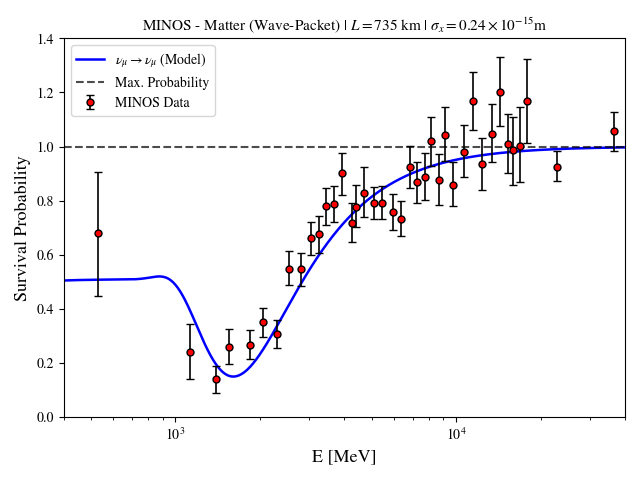}
    \caption{The three flavored Survival Probability of $\nu_{\mu}$ is plotted inspired from the wave packet approximation.}
    \label{fig:wave MINOS}
\end{figure}

\textit{Sequential removal.} We then removed bins one at a time, at each
step dropping the bin whose removal lowers $\Delta\chi^2$ most
(Fig.~\ref{fig:jackknife}). After one to five removals,
$\Delta\chi^2 = 12.4,\ 7.6,\ 4.1,\ 0.8$ and $0.3$. The five bins removed
are at $E\simeq1551,\ 1391,\ 528,\ 1837$ and $2059$~MeV. With these
removed, the plane-wave model describes the remaining 34 bins acceptably
($\chi^2_{\rm PW}=47.5$), and the fit shows no preference for a finite
$\sigma_x$.

\textit{Interpretation.} The preference for a finite wave-packet width
therefore arises from a localized cluster of five bins at
$0.5$--$2.1$~GeV, the same region in which the plane-wave prediction
visibly misses the data (Fig.\ref{fig:plane MINOS}), and not from the spectrum as a whole.
Because the sequential procedure removes the bins that carry the signal by
construction, it identifies where the preference originates; it does not
by itself show that the effect is absent. Since our fit uses only the
published statistical uncertainties, we cannot determine whether this
cluster reflects wave-packet decoherence or correlated systematic effects
in neighboring bins. We therefore do not quote a Gaussian-equivalent
significance, and treat $\sigma_x\approx0.24\times10^{-15}$~m as a
localized hint, adopted as a benchmark value for the predictions that
follow.

\begin{table}[t]
\caption{Greedy sequential removal of MINOS energy bins. At each step the
bin whose removal lowers $\Delta\chi^2=\chi^2_{\rm PW}-\chi^2_{\rm min}$
most is dropped and $\sigma_x$ is refit. For $n\ge6$ the minimum moves to
the upper edge of the scanned range ($3.16\times10^{-14}$~m), i.e. the
plane-wave limit.}
\label{tab:jackknife}
\begin{ruledtabular}
\begin{tabular}{cccc}
Bins removed & Bin removed (MeV) & $\Delta\chi^2$ & Best-fit $\sigma_x$ (m) \\
\hline
0 & --   & 22.16 & $2.41\times10^{-16}$ \\
1 & 1551 & 12.42 & $2.91\times10^{-16}$ \\
2 & 1391 & 7.55  & $1.56\times10^{-16}$ \\
3 & 528  & 4.05  & $1.56\times10^{-16}$ \\
4 & 1837 & 0.79  & $4.24\times10^{-16}$ \\
5 & 2059 & 0.31  & $5.86\times10^{-16}$ \\
\hline
6 & 1125 & 0.00  & $\to\infty$ (plane wave) \\
7 & 2546 & 0.00  & $\to\infty$ (plane wave) \\
8 & 6353 & 0.00  & $\to\infty$ (plane wave) \\
\end{tabular}
\end{ruledtabular}
\end{table}

\subsection{Explanation of the Wavepacket Model}
Possible explanation for such effect can be formed by defining the coherence and oscillation length - $L^{coh}$ and $L^{osc}$ for the atmospheric neutrinos, respectively.
\begin{itemize}
    \item For $E \approx 10 GeV$ : We evaluate using Eq.\ref{eq:Lcoh_matter} and Eq.\ref{eq:Losc_matter}, we get $L^{osc}_{31} \approx9,649.63km$ and $L^{coh}_{31} \approx 52,994 km$. Since, both $L^{coh}$ and $L^{osc}$ $\gg 735km$, which makes it nearly impossible for such baselines to observe the oscillation nature of neutrinos. \\
    Mathematically, 
    \[
P(\nu_\mu \rightarrow \nu_\mu) = 1 - 4\sum_{k>j} |W_{\mu k}|^2 |W_{\mu j}|^2 \sin^2\!\left(\frac{\Delta\lambda_{kj}L}{2}\right)
{e^{-(L/L_{kj}^{\mathrm{coh,m}})^2}}\] 
which in terms of Oscillation Length ($L^{osc,m}_{kj}$) can be written as :
\begin{equation}\label{Matterwave}
    P(\nu_\mu \rightarrow \nu_\mu) = 1 - 4\sum_{k>j} |W_{\mu k}|^2 |W_{\mu j}|^2 \sin^2\!\left(\pi\frac{L}{L^{osc,m}_{kj}}\right)
{e^{-(L/L_{kj}^{\mathrm{coh,m}})^2}}
\end{equation}
We can easily cancel out the exponential term due to the very large $L^{coh,m}$, as $L^{coh,m}_{31}=52994 km$, $L^{coh,m}_{21}\approx 76967km$ and $L^{coh,m}_{23}\approx 170140 km$ also the oscillation lengths are $L^{osc,m}_{31}\approx9649.63 km$, $L^{osc,m}_{21}\approx 14014.97km$ and $L^{osc,m}_{23}\approx30980.19$ which almost neglects out the sine terms in Equation \ref{Matterwave} as all the $L^{osc,m}_{kj}\gg$ baseline length of MINOS and hence, we get $P_{\nu_{\mu}\rightarrow\nu_{\mu}}\approx 1$, so for very high energy we would never observe oscillations within neutrino flavors. Moreover, when we neglected the exponential term, the probability expression turned out to be the plane wave approximation. 

\item For E $\approx$1.5 GeV : We solve it in the similar manner and in this case, we get $L^{coh}_{21} \approx 11430km$ and $L^{osc}_{21} \approx 13874.51km$, $L^{coh}_{31} \approx 1270.7km$ and $L^{osc}_{31} \approx 1542.43km$, $L^{coh}_{23} \approx 1429.6km$ and $L^{osc}_{23} \approx 1735.35km$. As we consider atmospheric neutrinos so, the main focus would be towards $L^{coh}_{31}$ and $L^{osc}_{31}$.\\
For range of energy $\approx (1000 - 1590)MeV$, as the $L^{coh}_{31}> L^{osc}_{31}$ and both the coherent and oscillation lengths are greater than the baseline of MINOS, the coherent length is twice as large as the oscillation length, which means that the mass eigenstates will always undergo interference and lead to flavor oscillations for MINOS experiment. Moreover, the baseline of MINOS  is half the size of $L^{osc}_{31}$ and since the oscillation length defines the spatial period over which a neutrino flavor transitions and oscillates \cite{Arguelles2023}, this increases the sine function in Eq.\ref{Matterwave} which reduces the survival probability and a drop is observed. \\

Alongside, here the phase ratios used in the exponential terms are not negligible unlike the previous case, 
\begin{equation}
\frac{L}{L^{osc,m}_{31}} \approx 0.4765, \qquad \frac{L}{L^{osc,m}_{21}} \approx 0.0529, \qquad \frac{L}{L^{osc,m}_{23}} \approx 0.4235
\end{equation}
This effect alters the oscillation profile, resulting in a curve that differs from the conventional plane-wave case and fits the realistic case.

\item For $E\approx 0.525$ GeV :  We observed that the oscillation nature of neutrinos explained via plane wave approximation doesn't give a proper explanation for such low energy ranges. 
The wavepacket model states that as $L^{osc}_{31}$ and $L^{coh}_{31}$ are $ 544.16 km$ and $158.41km$, respectively, and both of these lengths are less than the baseline length, so the oscillation nature suppresses and hence, this length omits out the oscillatory nature of the atmospheric neutrinos  and we observe a straight line within $\approx(400-600)MeV$. Moreover, as $L/L_{31}^{coh}\approx 4.64$ so the coherence completely gets lost, the wave packets become completely separated, suppressing the interference term,$L/L_{31}^{osc}=1.35$ this means that the neutrino has already passed through the first oscillation minimum and is recovering, these effects lead to the damping of neutrino oscillations.\\
Talking about the remaining phase ratio used in exponential terms  
\begin{equation}
\frac{L}{L^{osc,m}_{21}} \approx 0.0562, \qquad \frac{L}{L^{osc,m}_{23}} \approx 1.2945.
\end{equation}
These phase ratios used in the exponential terms gradually increase for low energy values and we don't observe an oscillation behavior for such low energy. 
\end{itemize}

The wavepacket approach therefore offers a useful framework for describing neutrino oscillations beyond the plane-wave approximation. A key 
parameter in this formalism is the wavepacket size $\sigma_{x}$:  we considered the values of $L^{osc}$ and $L^{coh}$ for energy ranges as explained above, depending on the MINOS data and set the value of $\sigma_{x} = 0.24 \times10^{-15} m $.
We further 
suggest that $\sigma_{x}$ be treated as a free phenomenological parameter --- 
analogous to the CP-violating phase $\delta_{\text{CP}}$ --- to be constrained 
directly from experimental data. By fitting $\sigma_{x}$ to the observed event 
rates at long-baseline experiments such as MINOS, one can obtain a physically 
motivated and data-driven determination of the wavepacket size, thereby improving 
the precision of theoretical predictions and providing a more faithful description 
of the observed oscillation pattern.

\section{DUNE Predictions via the Wave Packet Approach}\label{IV}
In this section, we evaluate the predictive framework of the Deep Underground Neutrino Experiment (DUNE) utilizing the wave packet treatment of neutrino oscillations as explained in Section \ref{Minos Wavepacket section}. 

DUNE is a next-generation, long-baseline neutrino experiment designed to resolve fundamental questions in neutrino physics. Operating at Fermilab, the experiment utilizes a high-intensity, accelerator-based neutrino beam produced by colliding high-energy protons into a graphite target. The resulting secondary mesons are magnetically focused and allowed to decay, generating a pure, wide-band muon neutrino (or antineutrino) flux. This beam is sampled first at a Near Detector complex on-site, before propagating 1,300 km through the Earth's crust to a Far Detector complex comprising massive Liquid Argon Time Projection Chambers (LArTPCs) situated at the Sanford Underground Research Facility.

While the primary baseline relies on this controlled accelerator mechanism, DUNE's deep underground location also grants it unique sensitivity to naturally occurring atmospheric neutrinos. Within this scope, we present a brief discussion on resolving the neutrino mass hierarchy—specifically examining how localized wave packet decoherence impacts these predictions when accounting for the broad energy spectra and extended path lengths characteristic of atmospheric neutrino fluxes \cite{EarthTomography,NunokawaCP}.

\subsection{Varying Earth - Matter Density}
\begin{figure}[t]
    \centering
\includegraphics[width=0.7\textwidth]{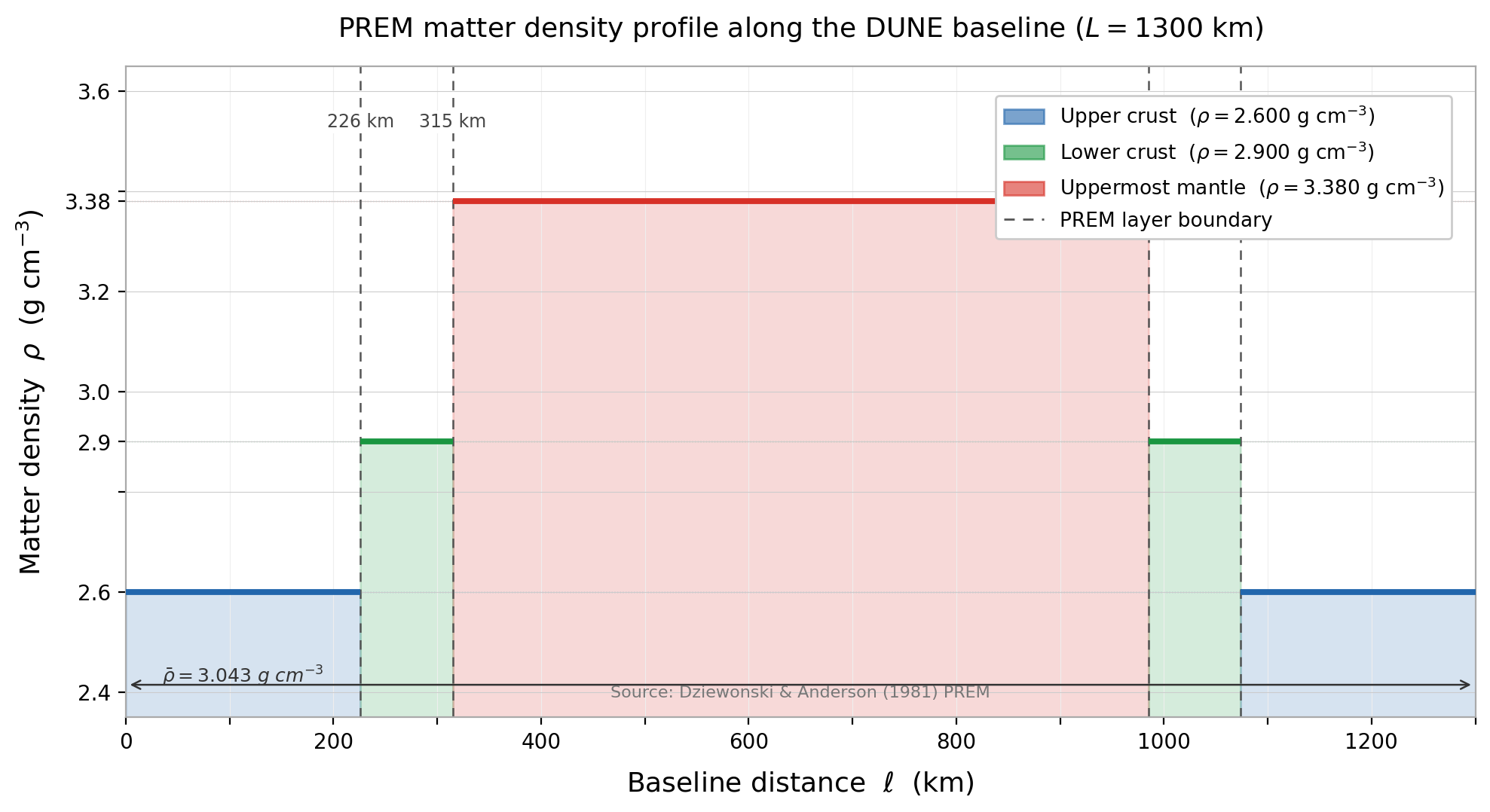}
    \caption{The plot shows the full PREM density profile  \cite{dziewonski1981preliminary} along the DUNE 1300 km baseline.The dashed vertical lines mark the four PREM layer boundary crossings at 226, 315, 985, and 1074 km. }
    \label{fig:DUNE PREM}
\end{figure}

As the baseline length of DUNE is 1300 km, so we can't neglect the varying matter effects unlike the MINOS. We referred \cite{dziewonski1981preliminary} to compute the varying matter density. A pictorial description of the varying density via the PREM model can be seen in Figure \ref{fig:DUNE PREM}.

\subsection{Wavepacket Model}
As discussed in Section \ref{Minos Wavepacket section}, the wavepacket treatment is more promising than widely used plane wave approximation, so we use the wavepacket model to the DUNE Experiment and perform a detailed study on it.\\

\begin{figure}[t] 
    \centering
\includegraphics[width=0.7\textwidth]{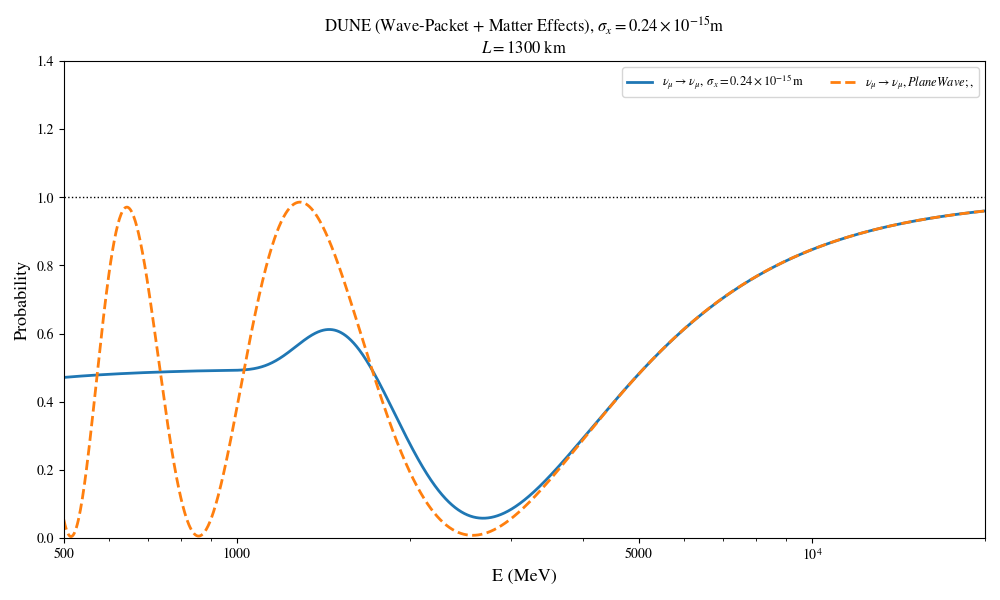}
    \caption{The three-flavor $\nu_\mu \rightarrow \nu_\mu$ 
survival probability at the DUNE baseline $L = 1300~\text{km}$ with $\sigma_x = 0.24 \times 10^{-15}~\text{m}$, computed under the wave packet (solid blue) and plane wave (dashed orange) approximations, incorporating varying Earth matter density via the PREM model. The energy range spans $0.5$--$20~\text{GeV}$.} 
    \label{fig: DUNE Wavepacket}
\end{figure}

Since the DUNE experiment employs the same accelerator-based neutrino production mechanism as MINOS, with neutrinos produced via pion and muon decay in the GeV energy range, the natural scale of the wavepacket size remains $\sigma_x \sim 10^{-15}~\text{m}$~\cite{Ettefaghi:2024}. We therefore adopt the same value $\sigma_x = 0.24 \times 10^{-15}~\text{m}$ 
for our DUNE predictions, consistent with the nuclear reaction scale of the production process.
The parameters used to compute our results for DUNE captured in Figure \ref{fig: DUNE Wavepacket} is taken from Table 2.

\begin{table}[t]\label{DUNE Parameter}
\centering
\begin{tabular}{|c|c|c|c|c|c|}
\hline
$\Delta m^2_{21}$ & $\Delta m^2_{31}$ & $\theta_{12}$ & $\theta_{13}$ 
& $\theta_{23}$ & $\delta_{CP}$ \\
\hline
$7.49 \times 10^{-5}\,\mathrm{eV}^2$ & $+2.513 \times 10^{-3}\,\mathrm{eV}^2$ 
& $33.68^\circ$ & $8.56^\circ$ & $43.3^\circ$ & $212^\circ$ \\
\hline
\end{tabular}
\caption{The standard three-flavor neutrino oscillation parameters utilized 
in the numerical analysis, corresponding to the best-fit points for Normal 
Ordering (NO) from the NuFIT~6.0 global analysis (IC24 with SK atmospheric 
data)~\cite{Esteban2024NuFit60}. The baseline is $L = 1300$~km.}
\label{tab:dune_params}
\end{table}

To understand the survival probability structure in Figure~4, we 
evaluate $L^{\text{osc,m}}_{31}$ and $L^{\text{coh,m}}_{31}$ at 
representative energies for the DUNE baseline $L = 1300$~km.

\begin{itemize}

\item For $E \approx 0.5$~GeV: We get $L^{\text{osc}}_{31} \approx 500.28$~km and $L^{\text{coh}}_{31} \approx 137.39$~km. Since both lengths are smaller than the DUNE baseline of 1300~km, with $L/L^{\text{coh}}_{31} \approx 9.46$ and $L/L^{\text{osc}}_{31} \approx 2.59$, the coherence is completely lost and the neutrino has passed through more than two full oscillation cycles. This results in a completely washed-out, flat survival probability in the low energy region, the oscillation suppression extends over a wider energy range than at MINOS.

\item For $E \approx 1.5$~GeV: We get $L^{\text{osc}}_{31} \approx 1503.67$~km and $L^{\text{coh}}_{31} \approx 1238.7$~km. Since $L^{\text{coh}}_{31}$ is less but $ \approx L $, coherence is lost but not much. However, $L/L^{\text{osc}}_{31} \approx 0.864$, so the baseline sits at 86.4\% of the first oscillation length, deeper than MINOS, pushing the survival probability minimum to a lower energy compared to MINOS. This explains the shifted position of the dip in Figure~4 relative to Figure~2.

\item For $E \gtrsim 5$~GeV: Both $L^{\text{osc}}_{31}$ and 
$L^{\text{coh}}_{31}$ grow large compared to 1300~km, the oscillation phase becomes negligible and $P(\nu_\mu \to \nu_\mu) \to 1$, with the wavepacket and plane wave predictions converging, consistent with Figure~4 at high energies.

\end{itemize}

\subsection{Mass Hierarchy for proposed DUNE model}
The neutrino mass ordering --- whether Normal Ordering (NO, $\Delta m^2_{31} > 0$, $m_1 < m_2 < m_3$) or Inverted Ordering (IO, $\Delta m^2_{31} < 0$, $m_3 < m_1 < m_2$) --- remains one of the central unresolved questions in neutrino physics, and its determination is a primary objective of DUNE~\cite{Esteban2024NuFit60}. In the wave packet formalism, the mass ordering enters through the matter-modified eigenvalue 
differences $\Delta\lambda_{kj}$, which govern both $L^{\mathrm{osc},m}_{kj}$ and $L^{\mathrm{coh},m}_{kj}$ via Eqs.~\eqref{eq:Lcoh_matter} and~\eqref{eq:Losc_matter}. 
The MSW potential $V_{cc}$ shifts the eigenvalues asymmetrically depending on the sign of $\Delta m^2_{31}$. 

The IO parameters are taken from the NuFIT~6.0 global analysis~\cite{Esteban2024NuFit60}.

\begin{table}[h]\label{DUNE Parameter (IO)}
\centering
\begin{tabular}{|c|c|c|c|c|c|}
\hline
$\Delta m^2_{21}$ & $\Delta m^2_{31}$ & $\theta_{12}$ & $\theta_{13}$ 
& $\theta_{23}$ & $\delta_{CP}$ \\
\hline
$7.49 \times 10^{-5}\,\mathrm{eV}^2$ & $-2.484 \times 10^{-3}\,\mathrm{eV}^2$ 
& $33.68^\circ$ & $8.59^\circ$ & $47.9^\circ$ & $274^\circ$ \\
\hline
\end{tabular}
\caption{The standard three-flavor neutrino oscillation parameters utilized 
in the numerical analysis, corresponding to the best-fit points for Inverted 
Ordering (IO) from the NuFIT~6.0 global analysis (IC24 with SK atmospheric 
data)~\cite{Esteban2024NuFit60}. The baseline is $L = 1300$~km.}
\label{tab:dune_params}
\end{table}

\begin{figure}[t]
    \centering
\includegraphics[width=0.7\textwidth]{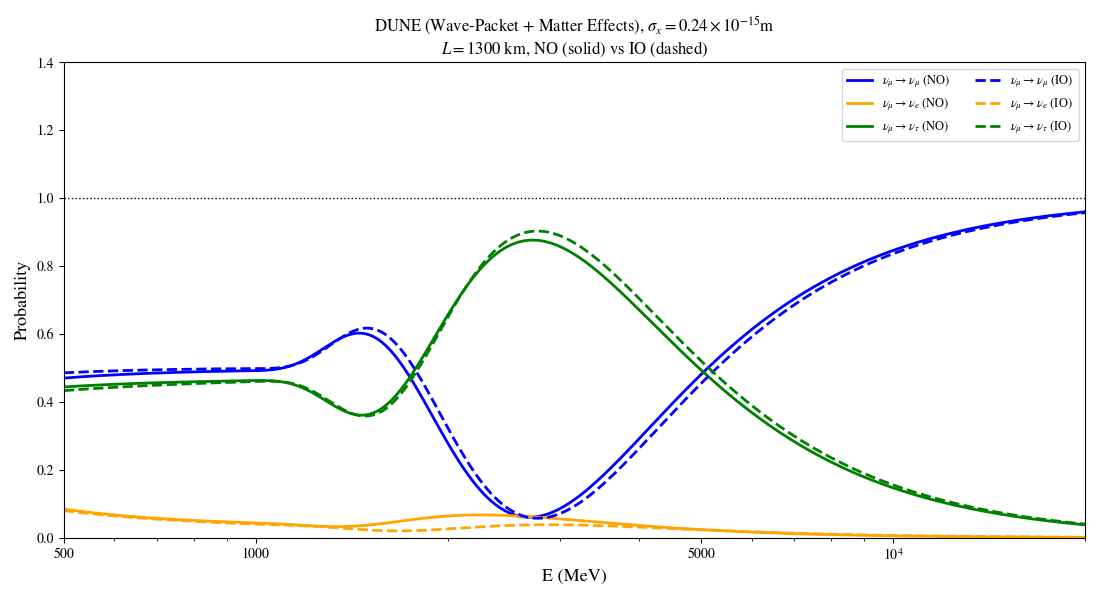}
    \caption{Three-flavor oscillation probabilities $P(\nu_\mu \to \nu_\alpha)$ ($\alpha = \mu, e, \tau$) at the DUNE baseline $L = 1300$~km with $\sigma_x = 0.24 \times 10^{-15}$~m under the wave packet approximation with varying PREM matter density, comparing Normal Ordering (NO, solid lines) and Inverted Ordering (IO, dashed lines) using NuFIT~6.0 IC24 with SK best-fit parameters~\cite{Esteban2024NuFit60}. The energy range spans $0.5$--$20$~GeV.}
    \label{fig:NO vs IO}
\end{figure}

Figures~7 show the three-flavor oscillation probabilities for NO and IO separately, while Figure~9 overlays both orderings directly for 
comparison. The overall oscillation structure is remarkably similar 
between NO and IO across the full energy range. Both orderings exhibit 
the same washed-out flat survival probability in the sub-GeV region 
($E \lesssim 800$~MeV) where coherence is completely lost, and both orderings
converge to $P(\nu_\mu \to \nu_\mu) \to 1$ at high energies 
($E \gtrsim 5$~GeV).

The most discernible differences appear in the second oscillation 
region around $E \approx 2$--$4$~GeV. The $\nu_\mu \to \nu_\tau$ 
transition (green) peaks slightly higher under IO ($\sim 0.93$) 
compared to NO ($\sim 0.90$), with the IO peak shifted toward 
higher energy. Correspondingly, the $\nu_\mu \to \nu_\mu$ survival 
minimum is marginally deeper and shifted to higher energy under IO. 
The $\nu_\mu \to \nu_e$ appearance channel (orange) shows a 
marginally higher probability under IO compared to NO in this region, 
though the difference remains small across the full spectrum.

These differences arise because the sign reversal of $\Delta m^2_{31}$ 
in IO modifies the matter eigenvalue differences $\Delta\lambda_{kj}$ 
asymmetrically through the MSW potential $V_{cc}$, shifting both 
$L^{\mathrm{osc},m}_{kj}$ and $L^{\mathrm{coh},m}_{kj}$ relative to 
their NO values. However, since the magnitude $|\Delta m^2_{31}|$ 
differs by only $\sim 1.2\%$ between NO and IO in the NuFIT~6.0 
best-fit, and since $L < L^{\mathrm{coh},m}_{31}$ for 
$E \gtrsim 3$~GeV in both orderings, the wave packet damping factor 
$e^{-(L/L^{\mathrm{coh},m}_{31})^2}$ remains close to unity for both 
orderings across the experimentally relevant DUNE beam energy range, 
leaving hierarchy discrimination primarily to the oscillation phases 
rather than the coherence structure.

\subsection{Oscillation Probability's Sensitivity by varying the range of $\sigma_x$ for DUNE}
\begin{figure}[t]
\includegraphics[width=1 \textwidth, height=0.4\textheight]{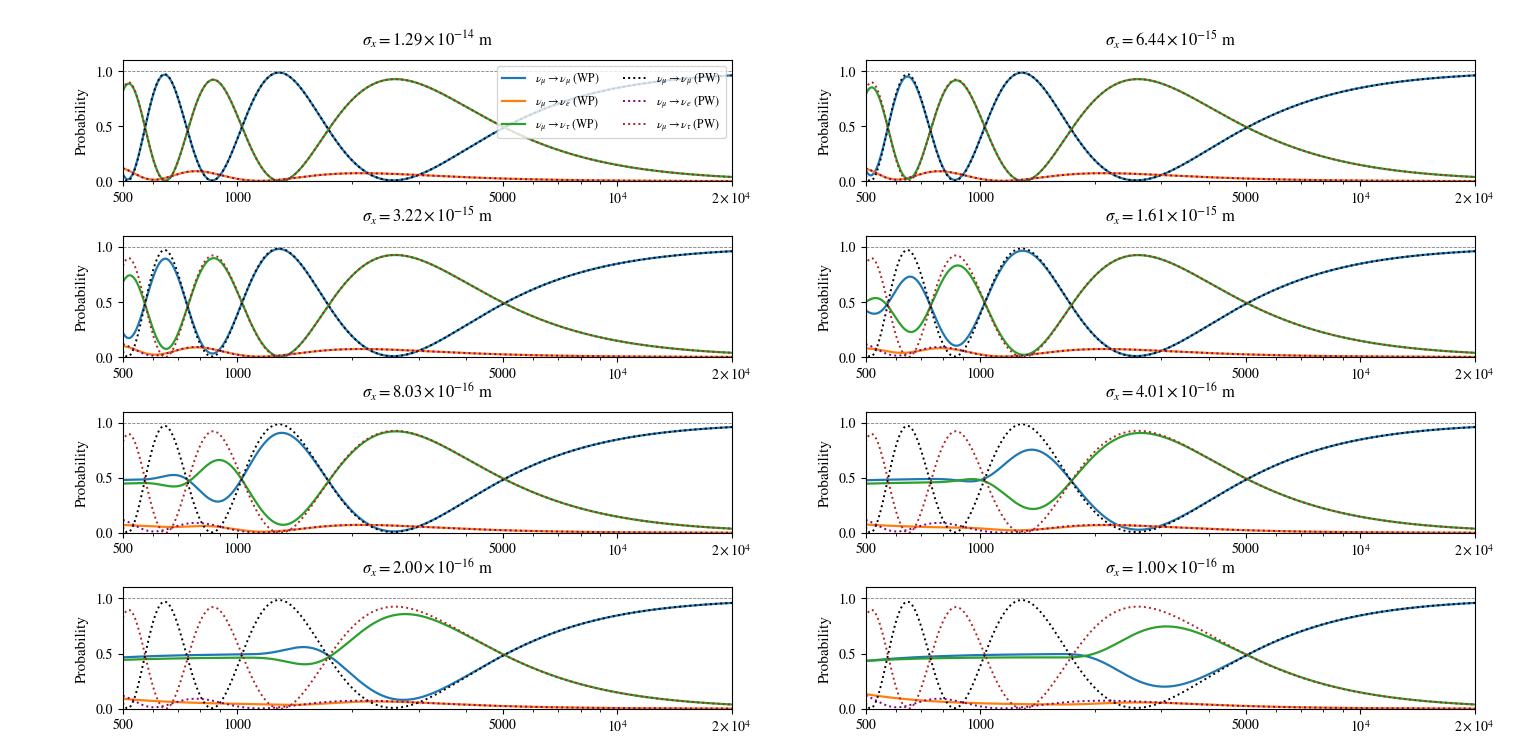}
    \caption{For the DUNE experiment's parameter we varied the values of $\sigma_x$ ranging from $1.29\times10^{-14}m$ to $10^{-16}m$. There are overall 8 probability vs Energy plots , each of them consists of 2 oscillation and 1 survival probability plots and the dotted plots represents the plane wave (PW) and the continuous plots represents the wave-packet plots (WP).}
    \label{fig:DUNE varying sigma}
\end{figure}

Now, since we do not have the data for DUNE, we would not constrain ourselves with $\sigma_{x}\approx10^{-15}m$, instead we would vary the value of $\sigma_x$ and check what results about the oscillation patterns can be drawn out of it.\\
We observe that for the DUNE's parameters, the value of $\sigma_x \leq 1.29\times10^{-14}m$ because if $\sigma_x$ gets higher than this value, the neutrino behaves like plane waves. 

\section{Classical Fisher Information: Sensitivity of $\sigma_x$} \label{CFI}
Recently, Górecki \cite{Gorecki:2023cac} made a detailed study about the Fisher Information studies. Several recent papers  \cite{chundawat2026leptoniccpphasedetermination,Dixit:2026yft,Farooq:2026eap,yadav2026quantumfisherinformationrevealing,Huang:2026bws} use the idea of Fisher Information -- classical (CFI) and 1uantum (QFI) --to study the sensitivity of unknown parameters such as $\delta_{CP},\theta_{12},\theta_{13},\theta_{23},etc.$ with a focus on both experimental and theoretical predictions. In this paper, we used the Classical Fisher Information (CFI) to study the sensitivity of $\sigma_x$ with respect to Neutrino Energy (E) for both the experiments - MINOS and DUNE. 

 Classical Fisher Information (CFI) provides measurement oriented information about the unknown parameter ($\lambda$), it tells how much information about $\lambda$ is associated with the measurement technique \cite{Farooq:2026eap}. Once a specific measurement strategy is adopted, such
as flavor detection at a far detector, the transition probabilities $P_{\alpha \rightarrow \beta}(\lambda,\Theta)$ define the CFI \cite{yadav2026quantumfisherinformationrevealing}. Mathematically, it is denoted by :
 \begin{equation}\label{5.1}
     F_C (\lambda;\Theta) = \sum_{\beta =e,\mu,\tau}\frac{1}{P_{\alpha \rightarrow \beta}(\lambda,\Theta)} \left(\frac{\partial P_{\alpha \rightarrow \beta}(\lambda,\Theta)}{\partial \lambda}\right)^2
 \end{equation}

Here, $F_C(\lambda;\Theta)$ means that we are studying the classical Fisher information of the unknown parameter $\lambda$ with respect to $\Theta$. $ P_{\alpha \rightarrow \beta}(\lambda,\Theta)$ means that the Probability is a function of both $\lambda$ and $\Theta$. For both the detectors, we consider $\alpha = \mu$

In our analysis, we employ the classical Fisher information $F_C(\sigma_x, E)$ to assess the
sensitivity of $\sigma_x$ as a function of neutrino energy $E$. As established in Section~3, the
plane-wave approximation fails to reproduce the MINOS data in the low-energy regime, where
wave-packet decoherence effects become significant. Consequently, if the peaks of
$F_C(\sigma_x, E)$ are found to lie within this low-energy range, this indicates that $\sigma_x$
is most precisely constrained precisely where the wave-packet formalism is most needed —
providing a direct, information-theoretic confirmation that $\sigma_x$ is well suited to
describing the low-energy behavior of the oscillation probability.\\

\noindent\textbf{Note:} We have constrained ourselves with $F_C(\sigma_x;E)$ and not the Quantum Fisher Information ($F_Q(\sigma_x;E)$), as QFI in this case deals with the maximum information of the sensitivity of the size of the wave-packet $\sigma_x$ with respect to the neutrino energy ($E$) for the neutrino states associated with the experiments (MINOS, DUNE), but we do not know if the experimental setup could capture the maximum information about $\ket{\nu_\alpha(t)}$. So, CFI, which is a function of probability and directly deals with the detection of the experiment, is more reliable for our analysis. Moreover, we have previously shown  the classical Fisher Information (CFI) is proportional to the $\frac{\partial^2\chi^2}{\partial \sigma_x^2}$ as proved in our previous work \cite{Farooq:2026eap}, establishing CFI as a statistically grounded sensitivity measure rather than a purely heuristic diagnostic. 

\subsection{$\sigma_x$ sensitivity for MINOS}

Figure~8 shows $F_C(\sigma_x, E)$ evaluated at the MINOS baseline $L = 735$~km for $\sigma_x = 0.24 \times 10^{-15}$~m. The $F_C(\lambda;\Theta)$ exhibits two prominent peaks, both located within the low-energy range $E \lesssim 1.5$~GeV — precisely the regime in which the plane-wave approximation was shown in Section~3 to disagree with the MINOS data. This shows that the sensitivity of $P_{\mu\mu}$ to $\sigma_x$ is concentrated exactly where the wave-packet formalism is required to correct the plane-wave prediction: for $\sigma_x = 0.24\times10^{-15}$~m, the Fisher information peaks in  the same neutrino energy region where the plane-wave approximation fails to fit the data points. At higher energies, $F_C(\sigma_x, E)$ drops sharply and remains negligible, consistent with the survival probability $P_{\mu\mu}$ (Figure~9b) approaching unity and losing sensitivity to $\sigma_x$ once the wave-packet and plane-wave predictions converge.

\begin{figure}[t]
    \centering
    \begin{subfigure}[b]{0.45\textwidth}
        \centering
        \includegraphics[width=\textwidth]{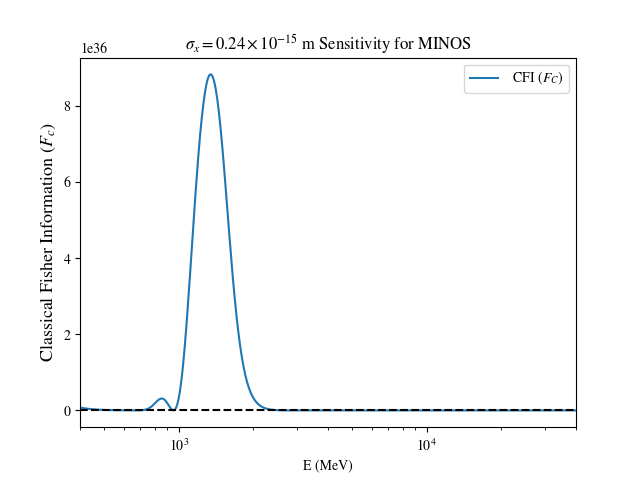}
        \caption{Classical Fisher information $F_C(\sigma_x, E)$ for $\sigma_x = 0.24\times10^{-15}$~m at the MINOS baseline, shown on its raw scale.}
        \label{9a}
    \end{subfigure}
    \hfill
    \begin{subfigure}[b]{0.45\textwidth}
        \centering
       \includegraphics[width=\textwidth]{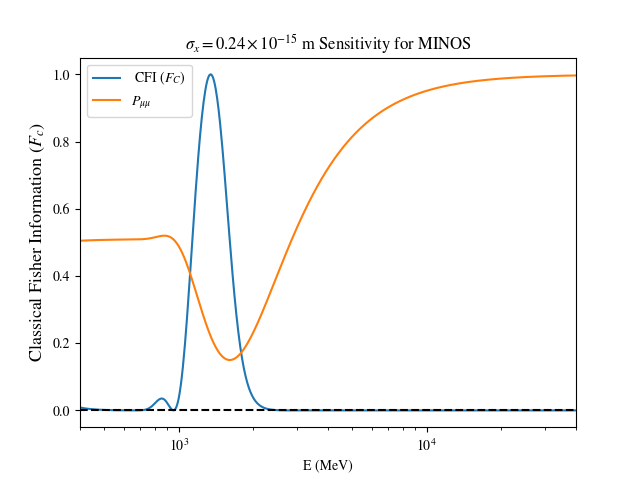}
        \caption{$F_C(\sigma_x, E)$ normalized to unit maximum, overlaid with the survival probability $P_{\mu\mu}$, to compare the location of the Fisher information peaks against the oscillation pattern.}
        \label{9b}
    \end{subfigure}
    \caption{Classical Fisher information sensitivity of $\sigma_x$ for the MINOS experiment at $\sigma_x = 0.24\times10^{-15}$~m. Panel (a) shows $F_C$ on its natural scale; panel (b) shows $F_C$ normalized so its peak structure can be directly compared against $P_{\mu\mu}(E)$. The peaks of $F_C(\sigma_x,E)$ lie within the low-energy region $E\lesssim1.5$~GeV, coinciding with the range where the plane-wave approximation was shown to disagree with the MINOS data, confirming that $\sigma_x$ is best constrained precisely where the wave-packet formalism is required.}
    \label{9}
\end{figure}

\subsection{$\sigma_x$ sensitivity prediction for DUNE}
Since no experimental data currently exist for DUNE, we cannot validate $\sigma_x$ against observed event rates as done for MINOS in Section~5.1. Instead, we treat $\sigma_x$ as a free parameter and compute $F_C(\sigma_x, E)$ across the representative range $\sigma_x \in [10^{-16},\,1.29\times10^{-14}]$~m identified in Section~4.5, in order to map where the DUNE energy spectrum would be most sensitive to each candidate value of $\sigma_x$. This provides a picture of the optimal energy window for probing $\sigma_x$ once DUNE data become available, and illustrates how the sensitivity profile itself shifts with the assumed wave-packet size.
Figures~8--10 show $F_C(\sigma_x, E)$ for three representative values of $\sigma_x$ spanning this range. As $\sigma_x$ decreases, the coherence length $L^{\mathrm{coh},m}_{kj}$ (Eq.~\eqref{eq:Lcoh}) shortens, so decoherence sets in at correspondingly higher energies, and the peak of $F_C(\sigma_x, E)$ shifts accordingly. In particular, for $\sigma_x = 1.29\times10^{-14}$~m the Fisher information peaks near $E \approx 0.5$~GeV (Figure~8), whereas for $\sigma_x = 1\times10^{-16}$~m the peak shifts to $E \approx 2.5$~GeV (Figure~10). This shows that the energy range over which DUNE would be most sensitive to $\sigma_x$ is itself a function of the (unknown) true wave-packet size, underscoring the value
of a broadband measurement covering the full $0.5$--$20$~GeV range for a model-independent determination of $\sigma_x$.

This energy-dependent shift also suggests a practical diagnostic once DUNE data become available: if the plane-wave prediction fails to fit the observed event rates in a specific neutrino energy range, $F_C(\sigma_x, E)$ can be used to identify which value of $\sigma_x$ would place peak sensitivity at that same energy range, thereby guiding the wave-packet fit toward the value most likely to resolve the discrepancy — the same procedure used for MINOS in Section~5.1. If the plane-wave approximation fails broadly across the low-energy region, this points toward a smaller wave-packet size, $\sigma_x \sim 10^{-16}$~m, since Figure~10 shows the Fisher information sensitivity shifted to higher energies in this regime. If instead the plane-wave approximation fits well overall but fails only within a narrower energy window, this points toward correspondingly increasing the order of magnitude of $\sigma_x$ to shift the
peak sensitivity into that window.
\begin{figure}[t]
    \centering
    \begin{subfigure}[b]{0.45\textwidth}
        \centering
\includegraphics[width=\textwidth]{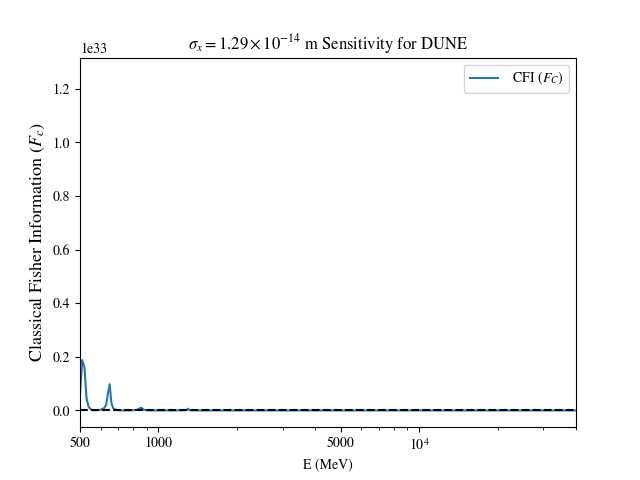}
    \caption{$F_C(\sigma_x,E)$ on its raw scale.}
        \label{fig:DUNE_CFI_1e-14_a}
    \end{subfigure}
    \hfill
    \begin{subfigure}[b]{0.43\textwidth}
        \centering
\includegraphics[width=\textwidth]{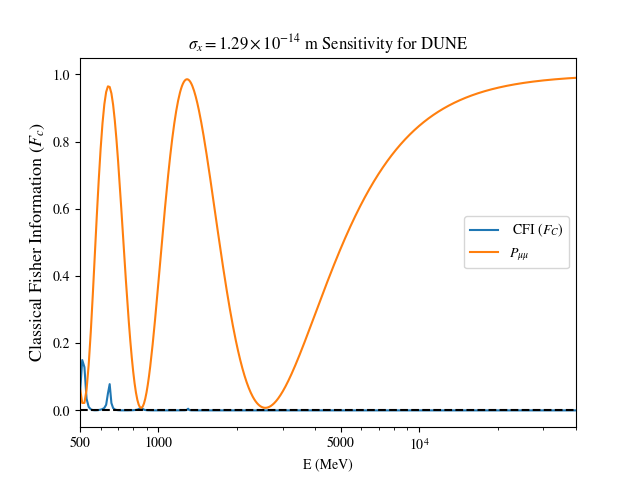}        \caption{$F_C(\sigma_x,E)$ normalized to unit maximum, overlaid with $P_{\mu\mu}(E)$.}
        \label{fig:DUNE_CFI_1e-14_b}
    \end{subfigure}
    \caption{Classical Fisher information sensitivity of $\sigma_x$ predicted for the DUNE
    baseline at $\sigma_x = 1.29\times10^{-14}$~m. Panel (a) shows $F_C(\sigma_x,E)$ on its
    natural scale; panel (b) shows $F_C(\sigma_x,E)$ normalized to unit maximum and overlaid
    with the predicted survival probability $P_{\mu\mu}(E)$, allowing the sensitivity peaks
    to be compared directly against the oscillation structure.}
\label{fig:DUNE_sigma_sensitivity_1e-14}
\end{figure}

\begin{figure}[t]
    \centering
    \begin{subfigure}[b]{0.43\textwidth}
        \centering
        \includegraphics[width=\textwidth]{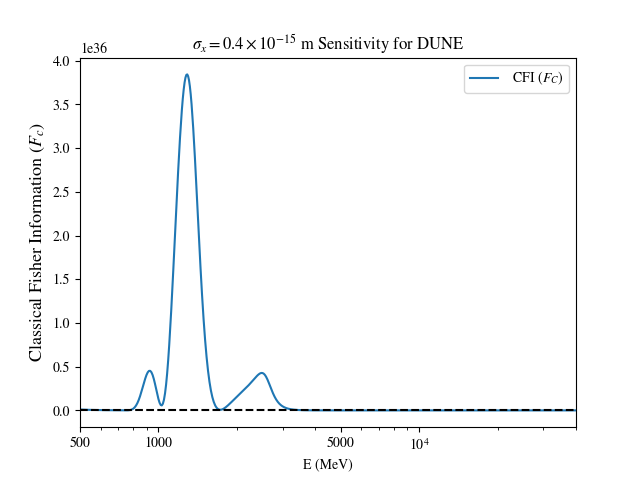}
        \caption{$F_C(\sigma_x,E)$ on its raw scale.}
        \label{fig:DUNE_CFI_0p4e-14_a}
    \end{subfigure}
    \hfill
    \begin{subfigure}[b]{0.43\textwidth}
        \centering
        \includegraphics[width=\textwidth]{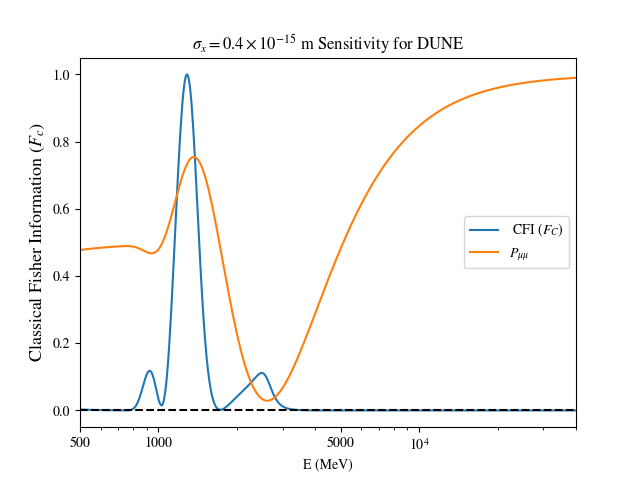}
        \caption{$F_C(\sigma_x,E)$ normalized to unit maximum, overlaid with $P_{\mu\mu}(E)$.}
        \label{fig:DUNE_CFI_0p4e-14_b}
    \end{subfigure}
    \caption{Classical Fisher information sensitivity of $\sigma_x$ predicted for the DUNE
    baseline at $\sigma_x \approx10^{-15}$~m. Panel (a) shows $F_C(\sigma_x,E)$ on its
    natural scale; panel (b) shows $F_C(\sigma_x,E)$ normalized to unit maximum and overlaid
    with the predicted survival probability $P_{\mu\mu}(E)$, allowing the sensitivity peaks
    to be compared directly against the oscillation structure.}
    \label{fig:DUNE_sigma_sensitivity_0p4e-14}
\end{figure}

\begin{figure}[H]
    \centering
    \begin{subfigure}[b]{0.45\textwidth}
        \centering
        \includegraphics[width=\textwidth]{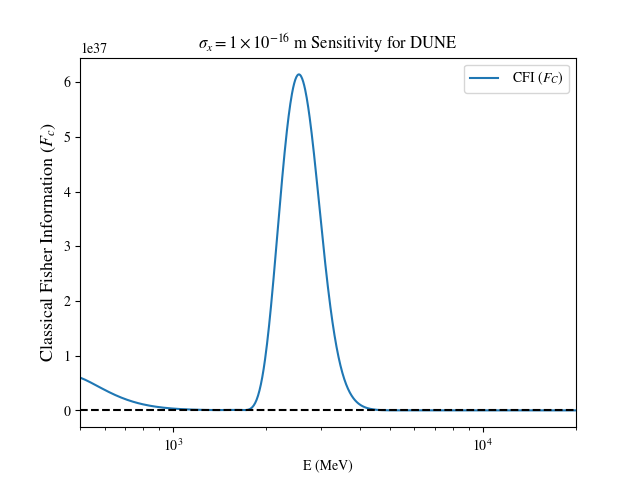}
        \caption{$F_C(\sigma_x,E)$ on its raw scale.}
        \label{fig:DUNE_CFI_1e-16_a}
    \end{subfigure}
    \hfill
    \begin{subfigure}[b]{0.45\textwidth}
        \centering
        \includegraphics[width=\textwidth]{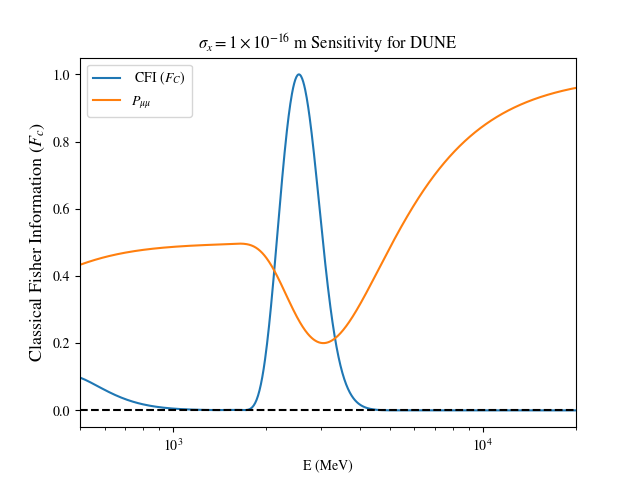}
        \caption{$F_C(\sigma_x,E)$ normalized to unit maximum, overlaid with $P_{\mu\mu}(E)$.}
        \label{fig:DUNE_CFI_1e-16_b}
    \end{subfigure}
    \caption{Classical Fisher information sensitivity of $\sigma_x$ predicted for the DUNE
    baseline at $\sigma_x = 1\times10^{-16}$~m. Panel (a) shows $F_C(\sigma_x,E)$ on its
    natural scale; panel (b) shows $F_C(\sigma_x,E)$ normalized to unit maximum and overlaid
    with the predicted survival probability $P_{\mu\mu}(E)$, allowing the sensitivity peaks
    to be compared directly against the oscillation structure. As $\sigma_x$ decreases across
    Figures~10--12, the coherence length $L^{\mathrm{coh},m}_{kj}$ shortens, shifting the
    energy at which $F_C(\sigma_x,E)$ peaks — narrowing and moving the region of maximal
    sensitivity to $\sigma_x$.}
    \label{fig:DUNE_sigma_sensitivity_1e-16}
\end{figure}

\section{Wave-particle Duality - Triality for MINOS and DUNE}\label{VI}

Recently, trade-off relations among local and non-local quantum resources have been investigated in the context of neutrino oscillations. In Refs.\cite{Bittencourt:2022tcl,Bittencourt:2023asd}, complete complementarity relations have been considered. In this line,  in Ref.\cite{Benerjee:2026}
neutrino oscillations were studied (in the plane wave approach) by means of the 
triality relation between path-predictability $\mathcal{P}^2$ , interference visibility $\mathcal{V}^2$ , and entanglement (I-concurrence) $\mathcal{E}^2$ for a general n-path interferometer, established as \cite{PhysRevA.105.032209} 
\begin{equation}
\mathcal{P}^2+\mathcal{V}^2+\mathcal{E}^2 = 1.
\end{equation}
In this framework, flavor conversion is treated as a which-path problem, with the detected flavor playing the role of the interfering ``quanton" and the undetected flavor sector playing the role of the path-detector it becomes entangled with.

\subsection{Neutrino States as Qubits : Density Matrix Formalism}
To apply the triality framework, we first cast the three-flavor neutrino
system as a multi-qubit quantum register. Each flavor mode
$\alpha \in \{e,\mu,\tau\}$ is treated as a two-level system (qubit), with
$|1\rangle_\alpha$ denoting ``a neutrino occupies flavor mode $\alpha$'' and
$|0\rangle_\alpha$ denoting ``it does not.'' The full system then lives in
the composite Hilbert space
\begin{equation}
\mathcal{H} = \mathcal{H}_e \otimes \mathcal{H}_\mu \otimes \mathcal{H}_\tau,
\qquad \dim\mathcal{H}=8,
\label{eq:hilbert-space}
\end{equation}
spanned by the computational basis
$\{|000\rangle,|001\rangle,|010\rangle,|011\rangle,|100\rangle,|101\rangle,|110\rangle,|111\rangle\}$,
where $|ijk\rangle \equiv |i\rangle_e\otimes|j\rangle_\mu\otimes|k\rangle_\tau$.

A single neutrino, by definition, occupies exactly one flavor mode at
production --- it cannot simultaneously ``be'' an electron-neutrino and a
muon-neutrino at $t=0$. This is a single-excitation constraint, structurally
identical to the exclusion that restricts a single fermion to occupying one
mode at a time. It confines the physically relevant flavor states to the
weight-one subspace of $\mathcal{H}$,
\begin{equation}
|\nu_e\rangle \equiv |100\rangle, \qquad
|\nu_\mu\rangle \equiv |010\rangle, \qquad
|\nu_\tau\rangle \equiv |001\rangle,
\label{eq:flavor-qubit-basis}
\end{equation}
with the remaining five basis states ($|000\rangle$, $|011\rangle$,
$|101\rangle$, $|110\rangle$, $|111\rangle$) carrying zero amplitude at all
times --- they are excluded by the same constraint, not merely unpopulated
by dynamics. This qubit encoding of flavor was introduced in Refs.\cite{Blasone:2007wp,Blasone:2007vw} and more recently used in Refs.\cite{Alok:2025qqr,SinghKoranga:2024rum}, and it is what makes the neutrino's flavor content a genuine multipartite quantum register rather than a single three-level system in disguise: the \emph{modes} are the qubits, and the flavor identity of the propagating neutrino is a coherent superposition across them.

\subsection*{Time evolution as an entangling process:}
A $\nu_\mu$ produced at $t=0$ starts in the product state
$|\nu_\mu(0)\rangle = |010\rangle = |0\rangle_e\otimes|1\rangle_\mu\otimes|0\rangle_\tau$
--- manifestly \emph{unentangled} across the three qubit modes, since each
factor is individually pure. Propagation, however, proceeds via the mass
eigenstates, which mix all three flavor modes according to the PMNS matrix.
Re-expressing the result back in the flavor basis in principle with the wavepacket formalism, gives
\begin{equation}
\rho_{e \mu \tau}^\mu(L) = \sum_{\beta\gamma} F_{\beta\gamma}^{\mu}(L) \left| \delta_{\beta e}\delta_{\beta \mu}\delta_{\beta \tau} \right\rangle \left\langle \delta_{\gamma e}\delta_{\gamma \mu}\delta_{\gamma \tau} \right| .
\qquad
F_{\beta\gamma}^{\mu}(L) = \sum_{jk} U_{\mu j}^* U_{\mu k} f_{jk}(L) U_{\beta j} U_{\gamma k}^*.
\label{eq:w-state-evolution}
\end{equation}
This is precisely a three-qubit W-state superposition --- a coherent sum of the three single-excitation basis kets, with time-dependent complex coefficients satisfying
$|F_{\tau \tau}^{\mu}(t)|^2+|F_{\mu \mu}^{\mu}(t)|^2+|F_{ee}^{\mu}(t)|^2=1$
By unitarity. So, now the density matrix formalism for $\ket{\nu_{\mu}(t)}$ is $\rho_{e\mu\tau}^{\mu}(t)$

\begin{equation}\label{rhomut}
\rho_{e\mu\tau}^{\mu}(t)
=
\begin{pmatrix}
    0 & 0 & 0 & 0 & 0 & 0 & 0 & 0\\
0 & F_{ee}^\mu & F_{e\mu}^\mu & 0 & F_{e\tau}^\mu & 0 & 0 & 0\\
0 & F_{\mu e}^\mu & F_{\mu\mu}^\mu & 0 & F_{\mu\tau}^\mu & 0 & 0 & 0\\
0 & 0 & 0 & 0 & 0 & 0 & 0 & 0\\
0 & F_{\tau e}^\mu & F_{\tau\mu}^\mu & 0 & F_{\tau\tau}^\mu & 0 & 0 & 0\\
0 & 0 & 0 & 0 & 0 & 0 & 0 & 0\\
0 & 0 & 0 & 0 & 0 & 0 & 0 & 0\\
0 & 0 & 0 & 0 & 0 & 0 & 0 & 0
\end{pmatrix}.
\end{equation}
We have adapted the same density matrix formalism as \cite{Benerjee:2026}. Now, we form the reduced density matrix assuming the detected flavor as $\ket{\nu_\mu(t)}$. So, we trace out the muon neutrino flavor \cite{Benerjee:2026}.
\begin{equation}
    \rho_{e\tau}^\mu = Tr_\mu (\rho^{\mu}_{e \mu \tau}) = 
    \begin{pmatrix}
F_{\mu\mu} & 0 & 0 & 0\\
0 & F_{ee}^\mu & F_{e\tau}^\mu & 0\\
0 & {F_{e\tau}^\mu}^{*} & F_{\tau\tau}^\mu & 0\\
0 & 0 & 0 & 0
\end{pmatrix}.
\end{equation}

Note that, in the wave-packet formalism adopted in this paper, the neutrino state is actually a mixed state, as a result of the time integration (see Appendix \ref{Appendix}).  However, the form of the density matrix \eqref{rhomut} is unchanged with respect to the pure case (see Ref.\cite{Bittencourt:2023asd}).

\subsection{Triality Relations: $\mathcal{P}^2,\mathcal{V}^2,\mathcal{E}^2$}
\subsubsection*{Path Predictability ($\mathcal{P}^2$) :}
 Predictability $\mathcal{P}^2$, quantifies how strongly a particular flavor outcome dominates over the others. In other words, it measures the degree of flavor imbalance, with larger values indicating a stronger preference for one flavor over the
other \cite{Benerjee:2026}. For an n-level system, predictability is defined as \cite{PhysRevA.105.032209}
\begin{equation}
    \mathcal{P}^{2}
=
\sum_{i=1}^{n}\rho_{ii}^{2}
-
\frac{1}{n-1}
\sum_{i\neq j}\rho_{ii}\rho_{jj}
\end{equation}
Here, where $\rho =\rho_{e\tau}^\mu(t)$ is the density matrix of the quanton and n = 4 (the dimension of the density matrix). The value of $\mathcal{P}^2$ varies from zero to one, so it is normalized. So, the final formula of predictability we get after using the reduced density matrix is :
\begin{equation}\label{6.8}    
\mathcal{P}^2=
\left(
P_{\mu\mu}^{2}
+
P_{\mu e}^{2}
+
P_{\mu\tau}^{2}
\right)
-
\frac{2}{3}
\left(
P_{\mu\mu}P_{\mu\tau}
+
P_{\mu\mu}P_{\mu e}
+
P_{\mu\tau}P_{\mu e}
\right).
\end{equation}
Within the overall triality relation, $\mathcal{P}^2$ represents the particle-like contribution associated with the detector
flavor. Its behavior therefore reflects the degree to which the detected neutrino maintains its identity during propagation, in
competition with the wave-like visibility and the nonlocal correlation term arising from entanglement with the remaining flavor
modes \cite{Benerjee:2026}.

\subsubsection*{Visibility ($\mathcal{V}^2$) :}
  Visibility also known as fringe contrast, so more is the value of $\mathcal{V}^2$, the sharper is the interference pattern formed \cite{PhysRevA.105.032209}, so it denotes the wave like characteristics for the neutrino state. It reflects the degree of quantum interference among different flavor modes and thus captures the wavelike nature of the neutrino oscillations; in the density-matrix formulation, it is governed by the off-diagonal elements of $\rho_{e\tau}^\mu (t)$, which encode the coherent superpositions responsible for flavor transitions \cite{Benerjee:2026}. For an n-level system, visibility is defined as \cite{PhysRevA.105.032209}
\begin{equation}\label{6.9}
    \mathcal{V}^2 \equiv \frac{n}{n-1} \sum_{i \neq j}|\rho_{ij}|^2
\end{equation}
where, $n = dim(\mathcal{H}_{e}\bigotimes\mathcal{H}_{ \tau}) = 4$, $\mathcal{H}_{e}$ and $\mathcal{H}_{\tau}$ is the Hilbert Space describing $\ket{\nu_e}$ and $\ket{\nu_{\tau}}$ states, respectively. So, finally in terms of probability we get :
\begin{equation}\label{6.10}
    \mathcal{V}^2 = \frac{8}{3}|F_{e\tau}^\mu|^2 
\end{equation}
From Eq.\ref{6.10} it is quite evident that the presence of coherent superposition is a necessary condition for observing interference effects or the wave nature of neutrinos.

\subsubsection*{Entanglement ($\mathcal{E}^2$) :}

Entanglement $\mathcal{E}^2$ quantifies the nonlocal correlation shared between the detected flavor $\nu_\mu$ and the remaining propagating subsystem $\{\nu_e,\nu_\tau\}$. Physically, it captures the portion of the neutrino's wave-like information that is not recoverable from the coherence term $\mathcal{V}^2$ alone, but is instead encoded in the correlations generated as the mass eigenstates mix during propagation \cite{Benerjee:2026}. For an $n$-level system, the generalized entanglement measure (I concurrence) is defined as \cite{PhysRevA.105.032209}
\begin{equation}
\mathcal{E}^2 = \frac{n}{n-1}\sum_{i\neq j}
\Big(\rho_{ii}\rho_{jj} - |\rho_{ij}|^2\Big),
\label{eq:E2-general}
\end{equation}
where, as before, $\rho = \rho^\mu_{e\tau}(t)$ and $n = \dim(\mathcal{H}_{e\tau}) = 4$.
Using the diagonal and off-diagonal elements of $\rho^\mu_{e\tau}$, we get entanglement as :
\begin{equation}
\mathcal{E}^2 = \frac{8}{3}\,\big(P_{\mu\mu}P_{\mu e}+P_{\mu e}P_{\mu \tau} + P_{\mu\mu}P_{\mu \tau} - |F_{e \tau}^\mu|^2\big).
\label{eq:E2-final}
\end{equation}
This expression shows that $\mathcal{E}^2$ is largest when the survival probability $P_{\mu\mu}$ is comparable in magnitude to the combined appearance probability $P_{\mu\tau}+P_{\mu e}$ --- i.e., precisely in the regime where the detected flavor is maximally undetermined between surviving and converting, and correspondingly smallest whenever either
$P_{\mu\mu}\to 0$ or $P_{\mu\tau}+P_{\mu e}\to 0$.

\paragraph{Verification of the triality identity.}
Adding Eqs.~\eqref{6.8}, \eqref{6.10}, and~\eqref{eq:E2-final},

\begin{align}
\mathcal{P}^2+\mathcal{V}^2+\mathcal{E}^2
&=
\Big(P_{\mu\mu}^2+P_{\mu e}^2+P_{\mu\tau}^2\Big)
-\frac{2}{3}
\Big(
P_{\mu\mu}P_{\mu\tau}
+P_{\mu\mu}P_{\mu e}
+P_{\mu\tau}P_{\mu e}
\Big)
\notag\\
&\quad
+\frac{8}{3}|F_{e\tau}^\mu|^2
+\frac{8}{3}
\Big(
P_{\mu e}P_{\mu\tau}
+P_{\mu\mu}P_{\mu\tau}
+P_{\mu\mu}P_{\mu e}
-|F_{e\tau}^\mu|^2
\Big)
\notag\\
&=
P_{\mu\mu}^2+P_{\mu e}^2+P_{\mu\tau}^2
+2P_{\mu\mu}P_{\mu\tau}
+2P_{\mu\mu}P_{\mu e}
+2P_{\mu\tau}P_{\mu e}
\notag\\
&=
\Big(P_{\mu\mu}+P_{\mu e}+P_{\mu\tau}\Big)^2.
\label{eq:triality-verify}
\end{align}
Using probability conservation, $P_{\mu\mu}+P_{\mu e}+P_{\mu\tau}=1$
the exact triality identity hence turns out to be :
\begin{equation}
\mathcal{P}^2+\mathcal{V}^2+\mathcal{E}^2 = 1
\label{eq:triality-final}
\end{equation}
In the next section, we  implement the above formalism in the study of neutrino oscillations in the framework of  experiments such as  MINOS and DUNE. Such a study is conducted within the wave-packet formalism: as stressed above, in this context the neutrino state is described by a mixed state. However, the above triality relation still holds \cite{Swain:2025otw}.

\subsection{Defining Triality for MINOS and DUNE}
In this section, we first focus on explaining the triality relations with respect to the energy bins for the MINOS and then we use the same ideology to explain the triality predictions for DUNE.\\

\subsubsection{Triality Analysis for MINOS ($\sigma_x \approx 10^{-15}m$) }

\begin{figure}[t]
\includegraphics[width=.7 \textwidth]{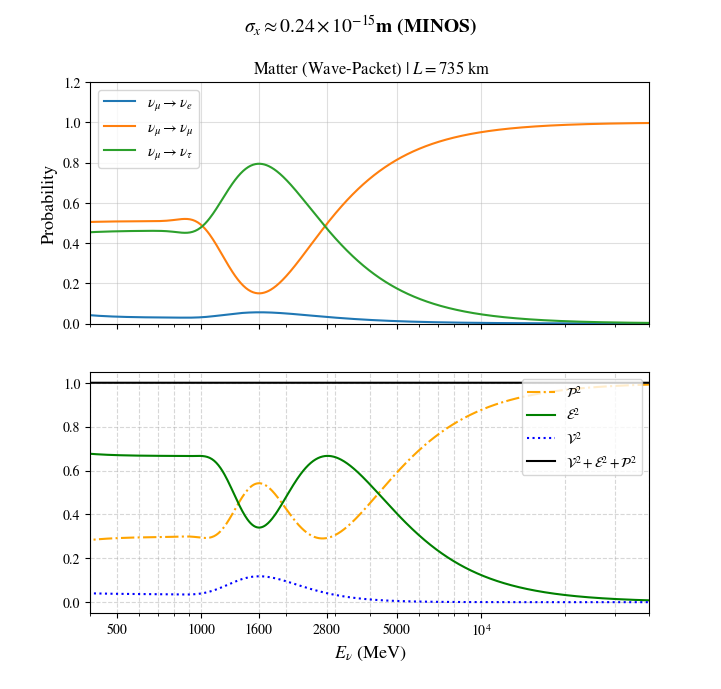}
   \caption{Neutrino oscillation probabilities for the MINOS, computed within the wave-packet formalism with a spatial coherence width $\sigma_x \approx 0.24 \times 10^{-15}$~m. \textit{Top panel:} Oscillation and survival probabilities as a function of neutrino energy $E_\nu$. \textit{Bottom panel:} Decomposition of the total transition probability into its coherence visibility ($\mathcal{V}^2$), entanglement ($\mathcal{E}^2$), and path perdictability ($\mathcal{P}^2$), with their sum - $\mathcal{V}^2 + \mathcal{E}^2 + \mathcal{P}^2 = 1$ shown as the black solid line.}
\label{fig:minos_wavepacket_prob}
\end{figure}

\textbf{For MINOS}, as shown in Figure \ref{fig:minos_wavepacket_prob}, We focus on 2 major energy bins $1.6$ GeV and $2.8$ GeV :
\subsubsection*{$E_\nu = 1.6$~GeV: Near the First Oscillation Extremum}

At this energy, the survival probability reaches a minimum, $P_{\mu\mu} \approx
0.150$, while the appearance probability $P_{\mu\tau}$ attains its maximum,
$P_{\mu\tau} \approx 0.791$, with $P_{\mu e} \approx 0.059$ remaining small but
non-negligible.

\begin{itemize}
    \item \textbf{Predictability $\mathcal{P}^2$ peaks.} One transition channel
    ($\nu_\mu \to \nu_\tau$) overwhelmingly dominates the flavor outcome. The
    neutrino's post-propagation identity is therefore highly determined:
    detection strongly favors a single outcome, corresponding to the
    particle-like regime of minimal flavor ambiguity.

    \item \textbf{Visibility $\mathcal{V}^2$ peaks.} From Eq.~(27), $\mathcal{V}^2 =
\tfrac{8}{3}|F^\mu_{e\tau}|^2$, where $F^\mu_{e\tau}$ is the coherence between
the two \emph{undetected} flavor modes surviving in the reduced density matrix
$\rho^\mu_{e\tau}$ (Eq.~(23)). This requires both undetected channels to be
populated \emph{and} to retain a definite relative phase. At $1.6$~GeV, $P_{\mu\tau}$ is large,
$P_{\mu e}$ is non-zero, and the wave-packet damping factors have not yet
washed out the oscillation phases, so the $e$--$\tau$ coherence is maximally
recoverable here.

    \item \textbf{Entanglement $\mathcal{E}^2$ reaches a minimum.} From
    Eq.~(29), $\mathcal{E}^2 = \tfrac{8}{3}(P_{\mu\mu}P_{\mu e} + P_{\mu\mu}
    P_{\mu\tau} + P_{\mu e}P_{\mu\tau} - |F^\mu_{e\tau}|^2)$. Since $P_{\mu\mu}$
    is small, the two terms involving it vanish almost entirely, and the
    remaining classical correlation $P_{\mu e}P_{\mu\tau}$ is largely offset by
    the sizable coherence term $|F^\mu_{e\tau}|^2$ already responsible for the
    $\mathcal{V}^2$ peak — the same coherence that shows up as observable
    interference in $\mathcal{V}^2$ is precisely what keeps $\mathcal{E}^2$
    suppressed here, since the two are built from the same $F^\mu_{e\tau}$ with
    opposite sign.
\end{itemize}

\textbf{Significance:} At $E_\nu = 1.6$~GeV, MINOS obtains the cleanest
simultaneous signature of both particle-like and wave-like neutrino behavior,
with predictability and visibility both near their peak values while
entanglement is strongly suppressed. Physically, this is the energy at which the
coherence between the undetected $\nu_e$ and $\nu_\tau$ modes is maximally
intact — the wave-packet decoherence factors have not yet washed out the phase
relationship encoded in $F^\mu_{e\tau}$ — so nearly all of the available
information is expressed either through outcome definiteness ($\mathcal{P}^2$)
or directly observable interference ($\mathcal{V}^2$), rather than being locked
away as entanglement with the undetected sector.

\subsubsection*{$E_\nu = 2.8$~GeV: Survival-Appearance Crossing}

At this energy, the survival and $\tau$-appearance probabilities become
approximately equal, $P_{\mu\mu} \approx P_{\mu\tau} \approx 0.5$, while
$P_{\mu e} \to 0$.

\begin{itemize}

     \item \textbf{Predictability $\mathcal{P}^2$ drops to an intermediate value
    ($\approx 0.33$).} With $P_{\mu\mu}$ and $P_{\mu\tau}$ now equally probable
    and $P_{\mu e} \to 0$, no single flavor outcome dominates as sharply as at
    $1.6$~GeV. The detected mode's identity is therefore only partially
    determined: the neutrino retains moderate particle-like character, but
    roughly two-thirds of the flavor information has shifted into the wave-like
    and entanglement contributions.

    \item \textbf{Visibility $\mathcal{V}^2 \approx 0.04$.} From Eq.~(27),
    $\mathcal{V}^2 = \tfrac{8}{3}|F^\mu_{e\tau}|^2$, which requires coherent
    population in \emph{both} undetected channels simultaneously. With
    $P_{\mu e} \to 0$, the $e$-channel amplitude vanishes, forcing
    $F^\mu_{e\tau} \to 0$ and hence $\mathcal{V}^2 \to 0$, regardless of the
    large $P_{\mu\tau}$.

    \item \textbf{Entanglement $\mathcal{E}^2$ reaches its maximum
    ($\approx 0.63$).} From Eq.~(29),
    \[
    \mathcal{E}^2 = \frac{8}{3}\Big(P_{\mu\mu}P_{\mu e} + P_{\mu\mu}P_{\mu\tau}
    + P_{\mu e}P_{\mu\tau} - |F^\mu_{e\tau}|^2\Big).
    \]
    With $P_{\mu e}\to 0$ (and hence $F^\mu_{e\tau}\to 0$ as above), the first
    and third terms drop out, leaving
    \[
    \mathcal{E}^2 \approx \frac{8}{3}\,P_{\mu\mu}P_{\mu\tau}
    \]
    consistent with the tabulated value. Physically, with no coherence term left
    to subtract off, this is the purely classical correlation between two
    comparably populated channels — the maximum such correlation can reach given
    $P_{\mu\mu}+P_{\mu\tau}\approx 1$.
\end{itemize}

\textbf{Significance:} The $2.8$~GeV bin highlights the physical distinction
between visibility and entanglement as independent carriers of wave-like
character. Although the directly observable interference between the undetected
channels has essentially vanished ($\mathcal{V}^2 \to 0$), the neutrino state is
not behaving as a separable, classical particle: it is maximally entangled with
the undetected sector ($\mathcal{E}^2 \approx 0.63$). This happens precisely
because the coherence term $|F^\mu_{e\tau}|^2$ that would otherwise be
subtracted from $\mathcal{E}^2$ has vanished along with $P_{\mu e}$, leaving the
full classical correlation between $P_{\mu\mu}$ and $P_{\mu\tau}$ exposed.
Visibility alone would therefore misleadingly suggest a predominantly
particle-like state at this energy, when in fact a substantial share of the
wave-like information has been redistributed into entanglement rather than
appearing as observable fringe-like coherence. This underscores the necessity of
the entanglement term $\mathcal{E}^2$ as an essential, non-optional component of
the triality relation.

\subsubsection{Triality Analysis for DUNE ($\sigma_x \approx 10^{-14}$~m)}
\begin{figure}[t]
\centering
\includegraphics[width=0.7 \textwidth]{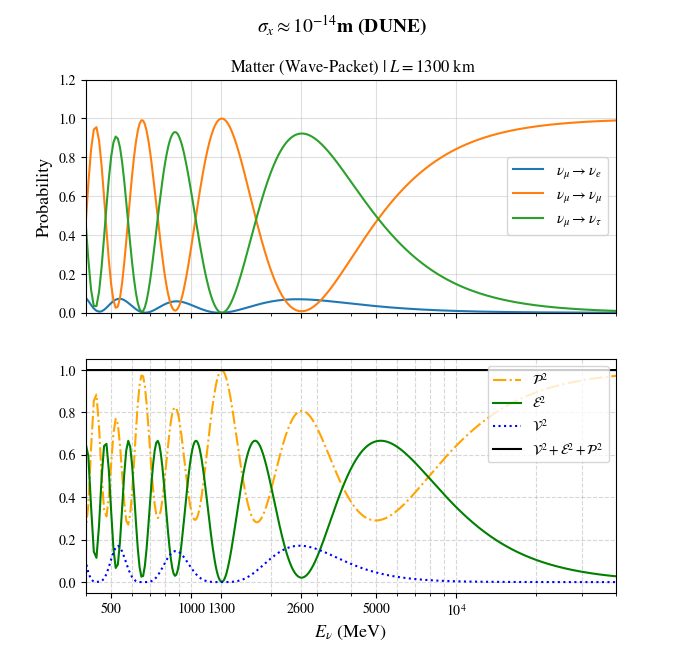}
    \caption{Neutrino oscillation probabilities for DUNE ($L = 1300$ km), computed within
the wave-packet formalism with a spatial coherence width $\sigma_x \approx 10^{-14}$ m.
\emph{Top panel:} Oscillation and survival probabilities as a function of neutrino
energy $E_\nu$. \emph{Bottom panel:} Decomposition of the total transition probability
into its coherence visibility ($\mathcal{V}^2$), entanglement ($\mathcal{E}^2$), and
path predictability ($\mathcal{P}^2$), with their sum
$\mathcal{V}^2 + \mathcal{E}^2 + \mathcal{P}^2 = 1$ shown as the black solid line.}
    \label{fig:CCR DUNE -14}
\end{figure}

For the DUNE baseline ($L = 1300$~km) with wave-packet coherence width
$\sigma_x \approx 10^{-14}$~m, Fig.~\ref{fig:CCR DUNE -14} displays the energy dependence of the triality components $\mathcal{P}^2$, $\mathcal{V}^2$, and $\mathcal{E}^2$. We focus on three representative energy bins, $E_\nu = 1.3$~GeV, $2.6$~GeV, and $5$~GeV, spanning the fully coherent,
matter-resonance, and mixed-decoherence regimes respectively.

\begin{table}[t]
\centering
\begin{tabular}{c c c c c c c}
\hline\hline
Experiment & $E$ (GeV) & $\mathcal{P}^2$ & $\mathcal{V}^2$ & $\mathcal{E}^2$ &
$P_{\mu\mu}$ & $P_{\mu e}$ \quad $P_{\mu\tau}$ \\
\hline
\multirow{2}{*}{MINOS ($\sigma_x\!\approx\!0.24\times10^{-15}$~m)}
 & 1.6 & {0.544} & {0.121} & {0.344} &
   {0.150} &{0.059} \quad {0.791} \\
 & 2.8 & {0.33} & {0.04} & {0.63} &
   {0.48} & {0.04} \quad {0.48} \\
\hline\hline
\end{tabular}
\caption{Triality condition parameters and oscillation probabilities for MINOS
($L = 735$~km) at $\sigma_x \approx 0.24\times10^{-15}$~m, evaluated at the two
characteristic energies $1.6$ and $2.8$~GeV.}
\label{tab:minos_triality}
\end{table}

\subsubsection*{$E_\nu = 1.3$~GeV: Full-Coherence Survival Peak}

At this energy the wave-packet separation remains small enough to sustain full
oscillatory coherence, and $E_\nu = 1.3$~GeV coincides with a low-energy survival
maximum:
\begin{align*}
    P_{\mu\mu} &\approx 1 \\
    P_{\mu e}  &\approx 0 \\
    P_{\mu\tau} &\approx 0
\end{align*}

\begin{itemize}
    \item \textbf{Predictability $\mathcal{P}^2 \approx 1$.} The neutrino
    survives as $\nu_\mu$ with near-certainty, so the flavor outcome is
    essentially fully determined: maximal particle-like character.

    \item \textbf{Visibility $\mathcal{V}^2 \approx 0$.} From Eq.~(27),
    $\mathcal{V}^2 = \tfrac{8}{3}|F^\mu_{e\tau}|^2$ requires coherent population
    in both undetected channels; with $P_{\mu e}, P_{\mu\tau} \approx 0$,
    $F^\mu_{e\tau}\to 0$ and visibility vanishes.

    \item \textbf{Entanglement $\mathcal{E}^2 \approx 0$.} From Eq.~(29), every
    term is a product involving $P_{\mu e}$ or $P_{\mu\tau}$; with both
    $\approx 0$, all terms vanish along with $\mathcal{E}^2$.
\end{itemize}

\textbf{Significance:} At $E_\nu = 1.3$~GeV the neutrino is in an almost fully
separable, single-path state, confirming that $\sigma_x \approx 10^{-14}$~m is
still narrow enough to preserve coherent oscillation at this baseline and
energy.

\subsubsection*{$E_\nu = 2.6$~GeV: Coherent Survival Regime with $\nu_e$-Appearance Onset}

Here the propagation sits close to a survival maximum, with a modest $\nu_e$-appearance
component switched on and $\tau$-appearance still suppressed:
\begin{align*}
    P_{\mu\mu} &\approx 0.927 \\
    P_{\mu e}  &\approx 0.073 \\
    P_{\mu\tau} &\approx 0
\end{align*}

\begin{itemize}
    \item \textbf{Predictability $\mathcal{P}^2 \approx 0.808$.} Predictability
    stays high because $P_{\mu\mu}$, not $P_{\mu\tau}$, dominates here — the
    small $P_{\mu e}$ is the only real competitor to survival.

    \item \textbf{Visibility $\mathcal{V}^2 \approx 0.165$.} Since $P_{\mu\tau}
\approx 0$, this value comes from $F^\mu_{e\tau}$ retaining coherence through
the mass-eigenstate cross terms even though the instantaneous
$\tau$-appearance probability itself is negligible. This is in fact the largest $\mathcal{V}^2$ among the three
representative bins, exceeding both the fully coherent $1.3$~GeV point
($\mathcal{V}^2\approx 0$) and the $5$~GeV crossover.

    \item \textbf{Entanglement $\mathcal{E}^2 \approx 0.027$.} With
    $P_{\mu\tau}\approx 0$, only the $P_{\mu\mu}P_{\mu e}$ term contributes
    appreciably, and it is largely offset by the same coherence term
    responsible for $\mathcal{V}^2$.
\end{itemize}

\textbf{Significance:} $E_\nu = 2.6$~GeV is where DUNE has maximal sensitivity
to $\nu_e$ appearance, and it is also where predictability and visibility are
jointly near their peak values while entanglement remains comparatively low —
a single-channel measurement here is least blind to the neutrino's full quantum
state.
\subsubsection*{$E_\nu = 5$~GeV: Onset of Matter-Driven Mixing}

By this energy the sharp low-energy oscillation structure has smoothed into a
broad, matter-modified transition region, with survival and $\tau$-appearance
probabilities becoming comparable:
\begin{align*}
    P_{\mu\mu} &\approx 0.48 \\
    P_{\mu e}  &\approx 0.04 \\
    P_{\mu\tau} &\approx 0.48
\end{align*}

\begin{itemize}
    \item \textbf{Predictability $\mathcal{P}^2 \approx 0.296$.} With two
    channels now comparably populated, no single outcome dominates, and the
    flavor identity is far less determined than at $1.3$ or $2.6$~GeV.

    \item \textbf{Visibility $\mathcal{V}^2 \approx 0.049$.} The small
    $P_{\mu e}$ limits the coherence amplitude $F^\mu_{e\tau}$ even though
    $P_{\mu\tau}$ is large, keeping $\mathcal{V}^2$ well below its $2.6$~GeV
    value despite the larger appearance probability here.

    \item \textbf{Entanglement $\mathcal{E}^2 \approx 0.655$.} From Eq.~(29),
    the dominant term is $\tfrac{8}{3}P_{\mu\mu}P_{\mu\tau} $, with the smaller $P_{\mu\mu} P_{\mu e}$ and $P_{\mu e}P_{\mu\tau}$ terms adding the rest; with
    $|F^\mu_{e\tau}|^2$ small (consistent with the small $\mathcal{V}^2$), there
    is little to subtract, so the classical correlation between the two
    comparably populated channels survives almost intact and $\mathcal{E}^2$
    dominates the triality budget.
\end{itemize}

\textbf{Significance:} At $E_\nu = 5$~GeV, entanglement — not predictability or
visibility — carries most of the neutrino's quantum information. This marks the
crossover region where survival and appearance probabilities converge, and the
resulting mixedness manifests predominantly as entanglement with the undetected
sector rather than as sharp predictability or directly observable interference,
illustrating why $\mathcal{V}^2$ alone would undersell the wave-nature present
here.
\begin{table}[t]
\centering
\begin{tabular}{c c c c c c c}
\hline\hline
Experiment & $E$ (GeV) & $\mathcal{P}^2$ & $\mathcal{V}^2$ & $\mathcal{E}^2$ &
$P_{\mu\mu}$ & $P_{\mu e}$ \quad $P_{\mu\tau}$ \\
\hline
\multirow{3}{*}{DUNE ($\sigma_x\!\approx\!10^{-14}$~m)}
 & 1.3 & {1} & {0} & {0} &
   {1} & {0} \quad  {0} \\
 & 2.6 & {0.808} & {0.165} & {0.027} &
   {0.927} & {0.073} \quad {0} \\
 & 5.0 & {0.296} & {0.049} & {0.655} &
   {0.48} & {0.04} \quad {0.48} \\
\hline\hline
\end{tabular}
\caption{Triality condition parameters and oscillation probabilities for DUNE at
$\sigma_x \approx 10^{-14}$~m, evaluated at the three characteristic energies
$1.3$, $2.6$, and $5$~GeV.}
\label{tab:dune_triality_1e-14}
\end{table}

\subsubsection{Triality Analysis for DUNE ($\sigma_x \approx 10^{-15}$~m)}

\begin{figure}[t]
\centering
\includegraphics[width=0.7\textwidth]{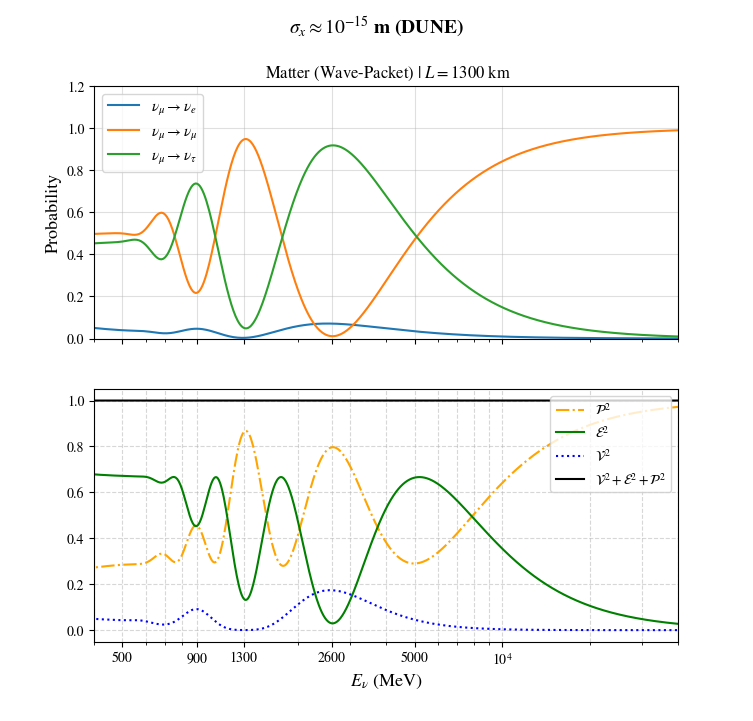}
    \caption{Neutrino oscillation probabilities for DUNE ($L = 1300$ km), computed within
the wave-packet formalism with a spatial coherence width $\sigma_x \approx 10^{-15}$ m.
\emph{Top panel:} Oscillation and survival probabilities as a function of neutrino
energy $E_\nu$. \emph{Bottom panel:} Decomposition of the total transition probability
into its coherence visibility ($\mathcal{V}^2$), entanglement ($\mathcal{E}^2$), and
path predictability ($\mathcal{P}^2$), with their sum
$\mathcal{V}^2 + \mathcal{E}^2 + \mathcal{P}^2 = 1$ shown as the black solid line.}
\label{fig:dune-triality-15}
    \label{fig:CCR DUNE -15}
\end{figure}

For the DUNE baseline ($L = 1300$~km) with wave-packet coherence width
$\sigma_x \approx 10^{-15}$~m, Fig.~\ref{fig:CCR DUNE -15} displays the energy
dependence of the triality components $\mathcal{P}^2$, $\mathcal{V}^2$, and
$\mathcal{E}^2$. We focus on three representative energy bins,
$E_\nu = 0.9$~GeV, $1.3$~GeV, and $2.6$~GeV.

\subsubsection*{$E_\nu = 0.9$~GeV: $\tau$-Appearance-Dominated, Predictability–Entanglement Near-Tie}

At this energy, the $\tau$-appearance probability is still rising toward its local
peak and already dominates the survival channel:

\begin{align*}
    P_{\mu\mu} &\approx 0.22  \\
    P_{\mu e}  &\approx 0.05 \\
    P_{\mu\tau} &\approx 0.73
\end{align*}

\begin{itemize}
    \item \textbf{Predictability $\mathcal{P}^2 \approx 0.45$.} With $P_{\mu\tau}$
    already the largest single contribution and climbing toward its local maximum,
    the flavor outcome is moderately well determined, though less sharply than at
    a fully separable point.

    \item \textbf{Visibility $\mathcal{V}^2 \approx 0.09$.} Since visibility
    depends on the coherence amplitude $|F^\mu_{e\tau}|^2$, the small $P_{\mu e}$
    keeps the observable $e$--$\tau$ interference weak even though $P_{\mu\tau}$
    itself is large: coherence needs \emph{both} undetected channels populated,
    and here only one of them is.

    \item \textbf{Entanglement $\mathcal{E}^2 \approx 0.46$.} With
    $\mathcal{E}^2$ built from the pairwise products
    $P_{\mu\mu}P_{\mu e} + P_{\mu e}P_{\mu\tau} + P_{\mu\mu}P_{\mu\tau}$ (minus the
    small coherence term), the sizable $P_{\mu\tau}$ together with non-negligible
    $P_{\mu\mu}$ and $P_{\mu e}$ combine to make entanglement almost equal to the predictibility
    of the three triality components at this bin. 
\end{itemize}

\textbf{Significance:} \textbf{Significance:} At $E_\nu = 0.9$~GeV, with $\nu_\mu \to \nu_\tau$ already the dominant channel, predictability and entanglement are nearly equal and jointly dominate the triality budget, while visibility is suppressed by the small $P_{\mu e}$. The near-equality $\mathcal{P}^2 \approx \mathcal{E}^2$ indicates that almost all of the coherence information available at this energy is being stored as entanglement with the undetected sector rather than surfacing as directly observable interference. This is a genuine crossover-type point well below the more familiar survival--appearance crossings (e.g.\ the MINOS 2.8~GeV bin or the 5~GeV DUNE crossover): here it is entanglement, not visibility, that
already carries the bulk of the wave-like character.
\subsubsection*{$E_\nu = 1.3$~GeV: Survival Probability Maximum}

At this energy the muon-neutrino survival probability reaches its maximum across
the three bins, with $\tau$-appearance and $e$-appearance both strongly
suppressed:

\begin{align*}
    P_{\mu\mu} &\approx 0.95 \\
    P_{\mu e}  &\approx 0 \\
    P_{\mu\tau} &\approx 0.05
\end{align*}

\begin{itemize}
    \item \textbf{Predictability $\mathcal{P}^2 \approx 0.86$.} With $P_{\mu\mu}$
    at its maximum and both appearance channels nearly vanishing, the flavor
    outcome is the most sharply determined of the three bins — $\mathcal{P}^2$
    peaks here precisely because $P_{\mu\mu}$ peaks here.

    \item \textbf{Visibility $\mathcal{V}^2 \approx 0$.} Visibility requires both
    undetected channels ($\nu_e$ and $\nu_\tau$) to be simultaneously populated;
    with $P_{\mu e} \approx 0$, the $e$--$\tau$ coherence term vanishes
    identically regardless of $P_{\mu\tau}$.

    \item \textbf{Entanglement $\mathcal{E}^2 \approx 0.14$.} From Eq.~(29),
    \[
    \mathcal{E}^2 = \frac{8}{3}\Big(P_{\mu\mu}P_{\mu e} + P_{\mu\mu}P_{\mu\tau}
    + P_{\mu e}P_{\mu\tau} - |F^\mu_{e\tau}|^2\Big).
    \]
    With $P_{\mu e} \approx 0$ (and hence $F^\mu_{e\tau}\approx 0$ as above), the
    first and third terms drop out, leaving
    \[
    \mathcal{E}^2 \approx \frac{8}{3}\,P_{\mu\mu}P_{\mu\tau}
    \]
    consistent with the tabulated $0.14$ up to the precision of the quoted
    $P_{\mu\mu}, P_{\mu\tau}$. Physically, this is the purely classical
    correlation between the surviving muon population and the small
    $\tau$-appearance leakage — with no coherence term to subtract off, it is
    the entire triality contribution left over once $\mathcal{V}^2$ vanishes.
\end{itemize}

\textbf{Significance:} $E_\nu = 1.3$~GeV is the point of maximum survival and
maximum predictability among the three bins: the neutrino is closest here to a
separable, particle-like state, with essentially all of the (small) remaining
triality budget carried by entanglement rather than visibility. However, unlike
at $\sigma_x \approx 10^{-14}$~m — where the analogous $1.3$~GeV bin has
$\mathcal{E}^2 \approx 0$ because $P_{\mu\tau}$ itself vanishes — here $P_{\mu\tau}$
is small but non-zero, so $\mathcal{E}^2 \approx \tfrac{8}{3}P_{\mu\mu}P_{\mu\tau}$
survives as a purely classical correlation term, with the coherence contribution
$|F^\mu_{e\tau}|^2$ switched off entirely by $P_{\mu e}\approx 0$. This is the
clean counterpoint to $E_\nu = 0.9$~GeV, where $\mathcal{P}^2 \approx \mathcal{E}^2$
and the coherent term contributes substantially to both $\mathcal{V}^2$ and
$\mathcal{E}^2$: the survival peak at $1.3$~GeV is where the state comes closest
to fully separable, but even here a narrower wave packet ($\sigma_x \approx
10^{-15}$~m rather than $10^{-14}$~m) leaves a small residue of classical
flavor correlation that the fully coherent limit does not show.
\subsubsection*{$E_\nu = 2.6$~GeV: Near-Maximal $\tau$-Appearance}

Here the propagation approaches the sharp $\tau$-appearance peak, with the muon
mode almost fully depleted:
\begin{align*}
    P_{\mu\mu} &\approx 0.015 \\
    P_{\mu e}  &\approx 0.078 \\
    P_{\mu\tau} &\approx 0.907
\end{align*}

\begin{itemize}
    \item \textbf{Predictability $\mathcal{P}^2 \approx 0.80$.} The
    $\nu_\mu \to \nu_\tau$ channel overwhelmingly dominates the flavor outcome,
    driving predictability close to its maximum — though still a bit below the
    $\mathcal{P}^2 \approx 0.86$ reached at the $1.3$~GeV survival peak, since
    $P_{\mu\tau}\approx 0.907$ here is slightly less dominant than $P_{\mu\mu}
    \approx 0.95$ was there.

    \item \textbf{Visibility $\mathcal{V}^2 \approx 0.17$.} From Eq.~(27),
    $\mathcal{V}^2 = \tfrac{8}{3}|F^\mu_{e\tau}|^2$ requires coherent amplitude
    in \emph{both} the $\nu_e$ and $\nu_\tau$ channels simultaneously. Unlike at
    $1.3$~GeV, $P_{\mu e}\approx 0.078$ is now non-zero, and together with the
    large $P_{\mu\tau}\approx 0.907$ this is enough to sustain a sizable
    coherence term $F^\mu_{e\tau}$ — the largest $\mathcal{V}^2$ among the three
    bins in this table.

    \item \textbf{Entanglement $\mathcal{E}^2 \approx 0.03$.} From Eq.~(29),
    \[
    \mathcal{E}^2 = \frac{8}{3}\Big(P_{\mu\mu}P_{\mu e} + P_{\mu\mu}P_{\mu\tau}
    + P_{\mu e}P_{\mu\tau} - |F^\mu_{e\tau}|^2\Big).
    \]
    Here all three pairwise products are individually small or moderate
    ($P_{\mu\mu}\approx 0.015$ suppresses the first two terms almost entirely,
    leaving mainly $P_{\mu e}P_{\mu\tau}\approx 0.071$), and this classical
    correlation is largely cancelled by the sizable coherence term
    $|F^\mu_{e\tau}|^2$ already responsible for the large $\mathcal{V}^2$ above.
    The near-vanishing of $P_{\mu\mu}$ is what drives $\mathcal{E}^2$ to its
    minimum across all three bins: with almost no surviving muon amplitude,
    there is little classical population left to correlate with the undetected
    sector, and what little classical correlation exists is offset by the
    coherent term.
\end{itemize}

\textbf{Significance:} $E_\nu = 2.6$~GeV is the cleanest window among the three:
predictability is high and visibility reaches its largest value in this set,
while entanglement is minimized. This is the direct opposite of $E_\nu =
0.9$~GeV, where $\mathcal{P}^2\approx\mathcal{E}^2$ and entanglement carries
most of the coherence information — here the same coherence, sourced by a
non-zero $P_{\mu e}$ acting together with a large $P_{\mu\tau}$, surfaces almost
entirely as directly observable interference ($\mathcal{V}^2$) rather than as a
classical correlation hidden in $\mathcal{E}^2$. It is also the direct
counterpoint to $E_\nu = 1.3$~GeV, where the vanishing $P_{\mu e}$ shut off
$\mathcal{V}^2$ entirely and left only a small classical residue in
$\mathcal{E}^2$: at $2.6$~GeV that same coherence channel is switched back on,
and it is $\mathcal{V}^2$, not $\mathcal{E}^2$, that absorbs it. Whether this
still lines up with the corresponding $2.6$~GeV behavior at $\sigma_x \approx
10^{-14}$~m is worth rechecking against that table before keeping any explicit
cross-$\sigma_x$ comparison sentence.
\begin{table}[t]
\centering
\begin{tabular}{c c c c c c c}
\hline\hline
Experiment & $E$ (GeV) & $\mathcal{P}^2$ & $\mathcal{V}^2$ & $\mathcal{E}^2$ &
$P_{\mu\mu}$ & $P_{\mu e}$ \quad $P_{\mu\tau}$ \\
\hline
\multirow{3}{*}{DUNE ($\sigma_x\!\approx\!10^{-15}$~m)}
 & 0.9 & 0.45 & 0.09  & 0.46 & 0.22 & 0.05 \quad 0.73 \\
 & 1.3 & 0.86 & 0 & 0.14 & 0.95 & 0 \quad 0.05 \\
 & 2.6 & 0.80  & 0.17 & 0.03 & 0.015  & 0.078  \quad 0.907  \\
\hline\hline
\end{tabular}
\caption{Triality condition parameters and oscillation probabilities for DUNE at
$\sigma_x \approx 10^{-15}$~m, evaluated at the three characteristic energies
$0.9, 1.3$ and $2.6$~GeV.}
\label{tab:dune_triality_1e-15}
\end{table}

\subsubsection{Triality for DUNE ($\sigma_x \approx 10^{-16}$ m)}

\begin{figure}[t]
\centering
\includegraphics[width=0.7 \textwidth]{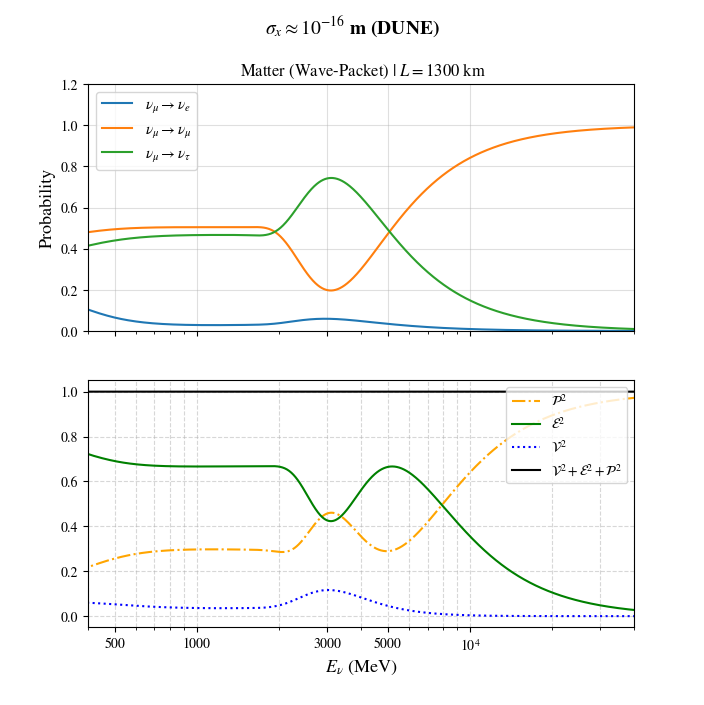}
    \caption{Neutrino oscillation probabilities for DUNE ($L=1300$ km), computed within
    the wave-packet formalism at a substantially reduced spatial coherence width
    $\sigma_x \approx 10^{-16}$ m. \emph{Top panel:} Oscillation and survival probabilities
    as a function of neutrino energy $E_\nu$. \emph{Bottom panel:} Decomposition of the
    total transition probability into path predictability ($\mathcal{P}^2$), visibility
    ($\mathcal{V}^2$), and entanglement ($\mathcal{E}^2$), with their sum
    $\mathcal{V}^2+\mathcal{E}^2+\mathcal{P}^2=1$ shown as the black solid line.}
    \label{fig:dune-triality-16}
\end{figure}

For the DUNE baseline ($L = 1300$ km) with a substantially narrower wave-packet coherence
width $\sigma_x \approx 10^{-16}$ m, Fig.~\ref{fig:dune-triality-16} displays the energy
dependence of the triality components $\mathcal{P}^2$, $\mathcal{V}^2$, and $\mathcal{E}^2$.
At this reduced value of $\sigma_x$, the coherence length $L_{kj}^{\mathrm{coh},m}$
(Eq.~\eqref{eq:Lcoh}) collapses well below the DUNE baseline over a much broader low-energy
window than in the $\sigma_x \approx 10^{-14}$ m or $10^{-15}$ m cases. As a direct
consequence, the entanglement term saturates at an almost energy-independent plateau,
$\mathcal{E}^2 \approx \mathrm{const.} \approx 0.67$, for $E_\nu \lesssim 2$ GeV — the
neutrino state is fully decohered into a maximally mixed superposition of the undetected
flavor modes well before any oscillatory structure can develop, and this plateau value is
set entirely by the asymptotic (fully decohered) mixing rather than by the instantaneous
oscillation phase. We focus on two representative energy bins beyond this plateau,
$E_\nu = 3$ GeV and $5$ GeV, which probe how the triality decomposition responds once
partial coherence is recovered.

\paragraph{$E_\nu = 3$ GeV: Partial Recoherence, $\tau$-Appearance Dominated.}
At this energy the wave packets have only partially recovered spatial overlap, and the
propagation probabilities are
\begin{equation}
P_{\mu\mu} \approx 0.2, \qquad P_{\mu e} \approx 0.063, \qquad P_{\mu\tau} \approx 0.737.
\end{equation}
\begin{itemize}
    \item \textbf{Predictability $\mathcal{P}^2 \approx 0.467$.} The $\nu_\mu \to \nu_\tau$
    channel dominates the flavor outcome by a wide margin, so the detected mode's identity
    remains fairly well determined despite the strong decoherence at lower energies.

    \item \textbf{Visibility $\mathcal{V}^2 \approx 0.105$.} With $P_{\mu e}$ small but
    non-zero and $P_{\mu\tau}$ large, the coherence amplitude $F^\mu_{e\tau}$ retains a
    modest but clearly non-vanishing value, consistent with Eq.~\eqref{6.9}.

    \item \textbf{Entanglement $\mathcal{E}^2 \approx 0.428$.} From Eq.~(29), the dominant
    contribution is $\tfrac{8}{3}P_{\mu\mu}P_{\mu\tau} \approx \tfrac{8}{3}(0.2)(0.737)
    \approx 0.393$, with the smaller $P_{\mu\mu}P_{\mu e}$ and $P_{\mu e}P_{\mu\tau}$ terms
    adding the rest and $|F^\mu_{e\tau}|^2$ (small, per the modest $\mathcal{V}^2$ above)
    subtracting little. The still-sizeable $P_{\mu\mu}$ combined with the large total
    appearance probability $P_{\mu e}+P_{\mu\tau}$ keeps the detected muon mode
    substantially correlated with the undetected sector, leaving entanglement comparable
    in magnitude to predictability itself.
\end{itemize}
\noindent\textbf{Significance:} At $E_\nu = 3$ GeV, predictability and entanglement
jointly carry most of the triality budget, with visibility a distinctly subleading
contribution. This is the energy at which the narrow wave packet first begins to exit
the fully decohered plateau, and the flavor content is simultaneously particle-like
(large $\mathcal{P}^2$) and strongly correlated with the undetected modes (large
$\mathcal{E}^2$), rather than displaying sharp observable interference.

\paragraph{$E_\nu = 5$ GeV: Survival--Appearance Crossover.}
By this energy the survival and $\tau$-appearance probabilities become comparable,
\begin{equation}
P_{\mu\mu} \approx 0.477, \qquad P_{\mu e} \approx 0.091, \qquad P_{\mu\tau} \approx 0.482.
\end{equation}
\begin{itemize}

    \item \textbf{Predictability $\mathcal{P}^2 \approx 0.292$.} With $P_{\mu\mu}$ and
    $P_{\mu\tau}$ nearly equal and no single channel dominating, predictability drops
    close to the intermediate values seen at the analogous crossover points for
    $\sigma_x \approx 10^{-14}$ m and $10^{-15}$ m (Tables~\ref{tab:dune_triality_1e-14} and
    \ref{tab:dune_triality_1e-15}).

    \item \textbf{Visibility $\mathcal{V}^2 \approx 0.039$.} The comparatively small
    $P_{\mu e}$ limits the coherence amplitude $F^\mu_{e\tau}$ despite the large
    $P_{\mu\tau}$, keeping visibility the smallest of the three triality components at
    this bin.

    \item \textbf{Entanglement $\mathcal{E}^2 \approx 0.669$.} From Eq.~(29), the
    dominant term is $\tfrac{8}{3}P_{\mu\mu}P_{\mu\tau}$ with the smaller $P_{\mu\mu}P_{\mu e}$ and $P_{\mu e}P_{\mu\tau}$ terms
    adding the rest and the small $|F^\mu_{e\tau}|^2$ (consistent with the small
    $\mathcal{V}^2$ above) subtracting little. With $P_{\mu\mu}$ and $P_{\mu\tau}$
    simultaneously large and closely matched, the classical correlation between the
    detected muon mode and the undetected tau mode survives almost intact, and
    entanglement dominates the triality decomposition almost entirely.
\end{itemize}
\noindent\textbf{Significance:} As at the crossover bins identified for
$\sigma_x \approx 10^{-14}$ m and $10^{-15}$ m, the $5$ GeV bin confirms that whenever
$P_{\mu\mu} \approx P_{\mu\tau}$, the resulting flavor mixedness is carried almost
entirely by entanglement with the undetected sector rather than by predictability or
directly observable interference. This behavior persists robustly across nearly two
orders of magnitude in $\sigma_x$, reinforcing that $\mathcal{E}^2$ is an essential,
non-optional component of the triality relation independent of the assumed wave-packet
size.

\begin{table}[htbp]
\centering
\begin{tabular}{c c c c c c c c}
\hline
Experiment & $E$ (GeV) & $\mathcal{P}^2$ & $\mathcal{V}^2$ & $\mathcal{E}^2$ &
$P_{\mu\mu}$ & $P_{\mu e}$ & $P_{\mu\tau}$ \\
\hline
\multirow{2}{*}{DUNE ($\sigma_x \approx 10^{-16}$ m)}
 & 3.0 & 0.467 & 0.105 & 0.428 & 0.200 & 0.063 & 0.737 \\
 & 5.0 & 0.292 & 0.039 & 0.669 & 0.477 & 0.091 & 0.482 \\
\hline
\end{tabular}
\caption{Triality condition parameters and oscillation probabilities for DUNE at
$\sigma_x \approx 10^{-16}$ m, evaluated at the two characteristic energies $3$ and
$5$ GeV.}
\label{tab:dune-16}
\end{table}

\section{Conclusion}\label{sec:conclusion}
 
In this work, we have developed a wave-packet treatment of three-flavor neutrino
oscillations that unifies two strands of inquiry usually pursued separately: the
kinematic question of how finite wave-packet coherence modifies the oscillation
probability itself, and the quantum-informational question of how the triality
relation $\mathcal{P}^2 + \mathcal{V}^2 + \mathcal{E}^2 = 1$ \cite{PhysRevA.105.032209}
is realized in a propagating neutrino state. Starting from a Gaussian wave-packet
ansatz for each mass eigenstate, we derived the full three-flavor transition
probability in matter, including explicit coherence-damping and localization-suppression
terms governed by the wave-packet width $\sigma_x$ (Eq.~\eqref{eq:prob_matter}).
Confronting this formalism with MINOS muon-neutrino disappearance data, we found that
a wave-packet width $\sigma_x \approx 0.24\times10^{-15}$~m resolves the well-known
failure of the plane-wave approximation in the low-energy region $E\lesssim1.5$~GeV
(Figs.~\ref{fig:plane MINOS} and \ref{fig:wave MINOS}), a regime in which the
coherence length $L^{\rm coh,m}_{31}$ and oscillation length $L^{\rm osc,m}_{31}$
(Eqs.~\eqref{eq:Lcoh_matter} and \eqref{eq:Losc_matter}) both become comparable to
or smaller than the MINOS baseline. Propagating the same formalism, with the Earth's
varying matter density incorporated via the PREM model, to the DUNE baseline
$L=1300$~km, we obtained wave-packet predictions for both the Normal and Inverted
mass orderings, and showed that hierarchy discrimination at DUNE is governed
primarily by the oscillation phases rather than by the coherence structure, since
the wave-packet damping factor $e^{-(L/L^{\rm coh,m}_{31})^2}$ remains close to
unity for $E\gtrsim3$~GeV under both orderings.
 
Using the classical Fisher information $F_C(\sigma_x,E)$, we further showed that
the sensitivity of the oscillation probability to $\sigma_x$ is concentrated
precisely in the low-energy region where the wave-packet and plane-wave predictions
diverge, both for MINOS at its fitted value $\sigma_x\approx0.24\times10^{-15}$~m
(Fig.~\ref{9}) and, predictively, for DUNE across the representative range
$\sigma_x\in[10^{-16},1.29\times10^{-14}]$~m (Figs.~\ref{fig:DUNE_sigma_sensitivity_1e-14},
\ref{fig:DUNE_sigma_sensitivity_0p4e-14}, and \ref{fig:DUNE_sigma_sensitivity_1e-16}).
This provides a direct, information-theoretic confirmation that $\sigma_x$ is best
constrained exactly where the wave-packet correction is physically required, and
offers a concrete diagnostic --- the energy at which $F_C(\sigma_x,E)$ peaks shifts
systematically with $\sigma_x$ --- that can guide a future wave-packet fit to DUNE
event rates once data become available.
 
The central novelty of this work lies in extending the triality decomposition of
Refs.~\cite{Benerjee:2026,PhysRevA.105.032209} --- originally formulated for
plane-wave neutrino oscillations and for photonic multipath interferometry,
respectively --- into the wave-packet regime, and in doing so for both MINOS and
DUNE across nearly two orders of magnitude in $\sigma_x$. Recasting the propagating
neutrino as a three-qubit flavor register and tracing out the detected $\nu_\mu$
mode, we verified the exact identity $\mathcal{P}^2+\mathcal{V}^2+\mathcal{E}^2=1$
at every energy and every value of $\sigma_x$ considered
(Fig.~\ref{fig:minos_wavepacket_prob} for MINOS and
Figs.~\ref{fig:CCR DUNE -14}, \ref{fig:dune-triality-15}, and
\ref{fig:dune-triality-16} for DUNE), confirming that the triality relation is not
an artifact of the plane-wave limit but a structural feature of the wave-packet
formalism itself. More significantly, we demonstrated that the wave-packet width
leaves an observable imprint on the \emph{internal redistribution} among
$\mathcal{P}^2$, $\mathcal{V}^2$, and $\mathcal{E}^2$ even in energy regions where
the survival probability $P_{\mu\mu}(E)$ itself is nearly indistinguishable across
different $\sigma_x$ choices. As $\sigma_x$ is reduced from $10^{-14}$~m to
$10^{-16}$~m, the low-energy entanglement plateau broadens and the crossover energy
at which $\mathcal{E}^2$ overtakes $\mathcal{P}^2$ as the dominant triality
component shifts accordingly (Tables~\ref{tab:dune_triality_1e-14},
\ref{tab:dune_triality_1e-15}, and \ref{tab:dune-16}), tracking the same
decoherence physics that governs $L^{\rm coh,m}_{kj}$. This shows that the triality
quantities constitute an independent, complementary probe of $\sigma_x$ beyond the
oscillation probability and the classical Fisher information alone --- precisely
the kind of imprint anticipated in Sec.~\ref{sec:intro}, where we noted that
$P_{\mu\mu}(E)$ and the wave-particle-entanglement quantities need not track the
same underlying coherence properties in the same way.
 
Across all cases examined, we find --- consistent with the pattern identified for
DUNE and T2K in the plane-wave analysis of Ref.~\cite{Benerjee:2026}
--- that whenever the survival and appearance probabilities become comparable,
$P_{\mu\mu}\approx P_{\mu\tau}$, the flavor mixedness is carried almost entirely by
entanglement with the undetected sector rather than by predictability or directly
observable interference, and that the cleanest simultaneous access to both
particle-like and wave-like neutrino character occurs where predictability and
visibility are jointly near their peak values while entanglement is suppressed
(Table~\ref{tab:minos_triality}). That this qualitative structure persists robustly
under wave-packet decoherence, and merely shifts in energy with $\sigma_x$ rather
than breaking down, underscores that the triality relation is a resilient
organizing principle for the quantum information content of long-baseline
oscillation experiments, independent of whether the underlying propagation is
treated as an idealized plane wave or as a localized wave packet of finite
coherence.
 
Taken together, our results argue that the wave-packet formalism, combined with
classical Fisher information and quantum-informational triality diagnostics,
provides a theoretically consistent and experimentally motivated framework for
probing quantum foundations in long-baseline neutrino oscillation experiments. A
natural extension of this work would be to repeat the triality analysis with
$\nu_e$ chosen as the detector flavor, offering complementary insight into the
appearance channel, and to combine the wave-packet CFI sensitivity map developed
here with an explicit quantum Fisher information bound, once a definite detection
strategy for DUNE atmospheric and beam neutrinos allows the saturability of that
bound to be assessed.
\section{ACKNOWLEDGMENTS}
We would like to thank  André de Gouvêa and   Christopher Ternes for their insightful comments and constructive suggestions on the wave-packet formulation presented in this work. 

\appendix

\section{Wave Packet Formalism} \label{Appendix}
\subsection{Flavor State as a Superposition of Wave Packets}
\label{subsec:flavor}

The width $\sigma_p$ is determined by the momentum 
uncertainty of the production process. Normalization 
is straightforwardly verified: $\int_{-\infty}^{\infty}
|\psi_j(p)|^2\,dp = 1$.\\

The coordinate-space wave function is obtained via 
the Fourier transform
\begin{equation}
    \psi_j(x,t) = \frac{1}{\sqrt{2\pi}}
    \int_{-\infty}^{\infty} dp\;\psi_j(p)\,
    e^{ipx - iE_j(p)t},
    \label{eq:FT}
\end{equation}
where $E_j(p) = \sqrt{p^2 + m_j^2}$ is the relativistic 
dispersion relation. Since the Gaussian 
(\ref{eq:momentum_wp}) is sharply peaked around $p_k$, 
we expand $E_j(p)$ to first order :
\begin{equation}
    E_j(p) \approx E_j + v_j(p - p_j),
    \label{eq:energy_expansion}
\end{equation}
where
\begin{equation}
    E_j \equiv E_j(p_j) = \sqrt{p_j^2 + m_j^2},
    \qquad
    v_j \equiv \left.\frac{\partial E_j}{\partial p}
    \right|_{p=p_j} = \frac{p_j}{E_j}
    \label{eq:group_velocity}
\end{equation}
are the mean energy and group velocity of the wave 
packet, respectively. On solving the Fourier transform, we get 
\begin{equation}
    \psi_j(x,t) = \left(2\pi\sigma_x^2\right)^{-1/4}
    \exp\!\left[-iE_j t + ip_j x 
    - \frac{(x - v_j t)^2}{4\sigma_x^2}\right],
    \label{eq:coord_wp}
\end{equation}
where
$\sigma_x = \frac{1}{2\sigma_p}$
is the spatial width of the wave packet. 
Equation~(\ref{eq:coord_wp}) describes a plane wave 
$e^{ip_k x - iE_k t}$ modulated by a Gaussian envelope 
of width $\sigma_x$ centered at $x = v_k t$, propagating 
at the group velocity $v_k$.

\subsection{Density Matrix and Time Average}
\label{subsec:density}

The pure state (\ref{eq:flavor_state}) defines the 
density matrix operator
\begin{equation}
    \hat{\rho}_\alpha(x,t) = 
    |\nu_\alpha(x,t)\rangle\langle\nu_\alpha(x,t)|.
    \label{eq:rho_pure}
\end{equation}

Since only the source-detector distance $L$ is 
accessible experimentally, we time-average 
$\hat{\rho}_\alpha(x,t)$ to obtain the stationary 
beam density matrix. 
\begin{equation}
    \rho_{\alpha}(x) = \int {dt \rho_\alpha{(x,t)}}
\end{equation}
Substituting (\ref{eq:coord_wp}) and on solving further we get 
\begin{align}
    \hat{\rho}_\alpha(x) = \sum_{k,j} 
    U^*_{\alpha k} U_{\alpha j}
    \exp\Bigg\{&-i\left[\frac{v_k + v_j}
    {v_k^2 + v_j^2}(E_k - E_j) 
    - (p_k - p_j)\right]x
    - \frac{(v_k - v_j)^2\,x^2}
    {4(v_k^2 + v_j^2)\sigma_x^2}
    - \frac{(E_k - E_j)^2}
    {4(v_k^2 + v_j^2)\sigma_p^2}
    \Bigg\}\,|\nu_k\rangle\langle\nu_j|.
    \label{eq:rho_stationary}
\end{align}

For ultra-relativistic neutrinos with $m_k \ll E$, 
we expand to leading order in $m_k^2/E^2$:
\begin{equation}
    E_k \simeq E + \xi_{\rm P}\frac{m_k^2}{2E},
    \qquad
    p_k \simeq E - (1-\xi_{\rm P})\frac{m_k^2}{2E},
    \qquad
    v_k \simeq 1 - \frac{m_k^2}{2E_k^2},
    \label{eq:relativistic}
\end{equation}
where $E$ is the neutrino energy in the zero-mass 
limit and $\xi_{\rm P}$ encodes the 
production kinematics~\cite{Giunti2004}. The $\xi_{\rm p}$  cannot be calculated in a quantum mechanical framework, but its value can be
estimated from energy-momentum conservation in the production process. We adopt $\xi_{\rm P} = 0$ throughout for simplicity, 
consistent with the kinematic constraints of pion and 
muon decay. Under (\ref{eq:relativistic}) and 
(\ref{eq:rho_stationary}), the time average $\rho_\alpha{(x)}$ reduces to
\begin{equation}
    \hat{\rho}_\alpha(x) = \sum_{k,j} 
    U^*_{\alpha k} U_{\alpha j}
    \exp\!\left[-i\frac{\Delta m^2_{kj}}{2E}\,x
    - \left(\frac{\Delta m^2_{kj}\,x}
{4\sqrt{2}\,E^2\,\sigma_x}\right)^{\!2} -\left(\xi_{P} \frac{\Delta m_{kj}^2}{4\sqrt{2} E \sigma_{p}^{ P2}}\right)^2\right]
    |\nu_k\rangle\langle\nu_j|,
    \label{eq:rho_relativistic}
\end{equation}
with $\Delta m^2_{kj} \equiv m_k^2 - m_j^2$.

\subsection{Transition Probability}
\label{subsec:probability}

The detection of flavor $\beta$ at position $x = L$ 
is described by an analogous detection operator 
$\hat{\mathcal{O}}_\beta(x-L)$ with detection 
momentum uncertainty $\sigma_p^{\rm D}$ and parameter 
$\xi_{\rm D}$. Evaluating the trace integral via Gaussian integration over $x$, the transition probability we get:
\begin{equation}
    P_{\nu_\alpha \to \nu_\beta}(L) = 
    \sum_{k,j} U^*_{\alpha k}\,U_{\alpha j}\,
    U_{\beta k}\,U^*_{\beta j}\;
    \exp\!\left[-2\pi i\frac{L}{L^{\rm osc}_{kj}}
    - \left(\frac{L}{L^{\rm coh}_{kj}}\right)^{\!2}
    - 2\pi^2\xi^2\!
    \left(\frac{\sigma_x}{L^{\rm osc}_{kj}}
    \right)^{\!2}\right],
    \label{eq:prob_full}
\end{equation}
where
\begin{align}
    L^{\rm osc}_{kj} &= \frac{4\pi E}
    {\Delta m^2_{kj}},
    \label{eq:Losc} \\[6pt]
    L^{\rm coh}_{kj} &= \frac{4\sqrt{2}\,E^2\,
    \sigma_x}{|\Delta m^2_{kj}|},
    \label{eq:Lcoh} \\[6pt]
    \sigma_x^2 &= \sigma_x^{{\rm P}2} 
    + \sigma_x^{{\rm D}2},
    \label{eq:sigma_combined} \\[6pt]
    \xi^2\sigma_x^2 &= \xi_{\rm P}^2\,
    \sigma_x^{{\rm P}2} + \xi_{\rm D}^2\,
    \sigma_x^{{\rm D}2}.
    \label{eq:xi_combined}
\end{align}
The three exponential terms in (\ref{eq:prob_full}) 
carry distinct physical content. The first is the 
standard oscillation phase, recovering the plane-wave 
result as $\sigma_x \to \infty$. The second is the 
coherence damping factor, suppressing oscillations 
when $L \gtrsim L^{\rm coh}_{kj}$ due to spatial 
separation of the wave packets. The third is the 
localization term, suppressing oscillations when 
the production or detection region is so sharply 
localized that mass eigenstates are kinematically 
resolvable. For $\xi_{\rm P} = \xi_{\rm D} = 0$, 
(\ref{eq:prob_full}) reduces to
\begin{equation}
    P_{\nu_\alpha \to \nu_\beta}(L) =
    \sum_{k,j} U^*_{\alpha k}\,U_{\alpha j}\,
    U_{\beta k}\,U^*_{\beta j}\;
    \exp\!\left[-2\pi i\frac{L}{L^{\rm osc}_{kj}}
    - \left(\frac{L}{L^{\rm coh}_{kj}}
    \right)^{\!2}\right].
    \label{eq:prob_xi0}
\end{equation}

\bibliographystyle{apsrev4-2}
\bibliography{LibraryNeutrinolug26}
\end{document}